\documentclass[11pt]{article}

\usepackage[utf8]{inputenc}
\usepackage[a4paper, margin=1.8cm, top=2cm, bottom =3cm]{geometry}

\usepackage{authblk}

\usepackage{mathtools}

\usepackage{amssymb}

\usepackage{dsfont}    

\usepackage{mathtools, nccmath,amsmath,amssymb,amsfonts,amstext, dsfont, color}
\usepackage{amsthm}

\usepackage{tikz}
\usetikzlibrary{automata,positioning,arrows}

\usepackage{booktabs}

\usepackage{subcaption}
\usepackage{mathtools, stmaryrd}
\usepackage{xparse} 
\usepackage{mathtools, stmaryrd}
\usepackage{algorithm}
\usepackage{algorithmicx}
\usepackage{algpseudocode}
\usepackage{nicefrac}
\usepackage{bm}
\usepackage{hyperref}

\usepackage{graphicx} 

\newtheorem{assumption}{Assumption}[section]

\newcommand{\Proba}[1]{\mathrm{Pr}\left(#1\right)}

\newcommand{\ProbaC}[2]{\mathrm{Pr}_{#1}\left(#2\right)}
\newcommand{\ProbaS}[2]{\mathrm{Pr}\left(#1\mid #2\right)}
\newcommand{\E}[1]{\mathbb{E}\left[#1\right]}
\newcommand{\Esub}[2]{\mathbb{E}_{#1}\left[#2\right]}

\newcommand{\Err}{\mathrm{Err}}
\newcommand{\CM}{\texttt{CM}}

\newcommand{\ElaSk}{\texttt{Elastic-Sketch}}
\newcommand{\aErr}{\overline{\mathrm{Err}}_{(0)}}

\DeclareMathOperator*{\argmax}{arg\,max}

\def\b0{\mathbf{0}}

\def\cB{\mathcal{B}}
\def\cI{\mathcal{I}}
\def\cA{\mathcal{A}}

\def\cO{\mathcal{O}}

\def\cI{\mathcal{I}}

\DeclarePairedDelimiter\floor{\lfloor}{\rfloor}

\usepackage{xcolor}

\definecolor{lightgray}{gray}{0.9}

\usepackage{mdframed} 

\newmdtheoremenv[
  linecolor=black,
  backgroundcolor=lightgray,
  linewidth=1pt,
  innerleftmargin=5pt,
  innerrightmargin=5pt,
  innertopmargin=1pt,
  innerbottommargin=1pt
]{theoremSp}{Theorem}

\newmdtheoremenv[
  linecolor=black,
  backgroundcolor=lightgray,
  linewidth=1pt,
  innerleftmargin=5pt,
  innerrightmargin=5pt,
  innertopmargin=1pt,
  innerbottommargin=1pt
]{lemmaSp}{Lemma}

\newmdtheoremenv[
  linecolor=black,
  backgroundcolor=lightgray,
  linewidth=1pt,
  innerleftmargin=5pt,
  innerrightmargin=5pt,
  innertopmargin=1pt,
  innerbottommargin=1pt
]{corollarySp}{Corollary}

\newmdtheoremenv[
  linecolor=black,
  backgroundcolor=lightgray,
  linewidth=1pt,
  innerleftmargin=5pt,
  innerrightmargin=5pt,
  innertopmargin=1pt,
  innerbottommargin=1pt
]{claimSp}{Claim}

\title{Elastic Sketch under Random Stationary Streams:\\ 
Limiting Behavior and Near-Optimal Configuration}
\author[1]{Younes Ben Mazziane}
\author[2]{Vinay Kumar B. R.}
\author[3]{Othmane Marfoq}

\affil[1]{University of Avignon, LIA, Avignon, France\\
\texttt{younes.ben-mazziane@univ-avignon.fr}}

\affil[2]{IIT Bombay, Mumbai, India\\
\texttt{vinaykumar.br@iitb.ac.in}}

\affil[3]{Meta, New York, USA\\
\texttt{omarfoq@meta.com}}

\date{}

\begin{document}

\maketitle

\begin{abstract}
   
\texttt{Elastic-Sketch} is a hash-based data structure for counting item's appearances in a data stream, and it has been empirically shown to achieve a better memory-accuracy trade-off compared to classical methods. This algorithm combines a \textit{heavy block}, which aims to maintain exact counts for a small set of dynamically \textit{elected} items, with a light block that implements \texttt{Count-Min} \texttt{Sketch} (\texttt{CM}) for summarizing the remaining traffic. The heavy block dynamics are governed by a hash function~$\beta$ that hashes items into~$m_1$ buckets, and an \textit{eviction threshold}~$\lambda$, which controls how easily an elected item can be replaced. We show that the performance of \texttt{Elastic-Sketch} strongly depends on the stream characteristics and the choice of~$\lambda$. Since optimal parameter choices depend on unknown stream properties, we analyze \texttt{Elastic-Sketch} under a \textit{stationary random stream} model---a common assumption that captures the statistical regularities observed in real workloads. Formally, as the stream length goes to infinity, we derive closed-form expressions for the limiting distribution of the counters and the resulting expected counting error. These expressions are efficiently computable, enabling practical grid-based tuning of the heavy and \texttt{CM} blocks memory split (via $m_1$) and the eviction threshold~$\lambda$. We further characterize the structure of the optimal eviction threshold, substantially reducing the search space and showing how this threshold depends on the arrival distribution. Extensive numerical simulations validate our asymptotic results on finite streams from the Zipf distribution.



\end{abstract}

\section{Introduction} 


Streaming algorithms process massive amounts of rapidly arriving data under strict resource constraints. They must support very fast per-update processing while using memory sublinear in the stream length, and thus return approximate answers~\cite{nelson2012sketching,LossyCounting_VLDB02,CHARIKAR2004CountSketch,ben2016heavy}. Such requirements arise routinely in network monitoring and clickstream analytics~\cite{Basat2021Salsa,AnomalyDetectionCMS2021}, and underpin large-scale data-processing systems. A common task is the detection of~$\phi$-heavy hitters, i.e., items whose frequency exceeds a fraction~$\phi$ of the total mass in a given time window. Under the streaming model above, this task typically involves approximate counting using either counter-based methods or sketch-based methods. 
Counter-based methods, such as \texttt{Misra-Gries}/\texttt{Frequent} \cite{Misra_Gries_82,Esa_Frequent02,Papadimitrou_Frequent03}, and \texttt{Space Saving} \cite{spaceSavingMetwally05}, maintain explicit counts for a dynamically chosen subset of items. On the other hand, sketch-based methods such as \texttt{Count-Sketch} \cite{CHARIKAR2004CountSketch} and \texttt{Count-Min} \textit{Sketch} ($\CM$) \cite{CORMODE2005CMS} update a compact hash-indexed array and provide fast estimates for any queried item. Sketches can be viewed as hash-based random projections of the high-dimensional frequency vector, with accuracy governed by the size of the hash-indexed array.


\texttt{Elastic Sketch}~\cite{ElasticSketch2019} combines counter-based and sketch-based ideas and exhibits a strong empirical memory-accuracy trade-off across a range of streaming tasks. It splits the available memory into two blocks: a \textit{heavy} block that aims to filter and explicitly count popular items, and a \textit{light} \texttt{Count-Min Sketch} (\texttt{CM}) block that summarizes the remaining traffic. In the heavy block, a hash function~$\beta$ maps items to~$m_1$ buckets, each storing and monitoring a single \textit{elected} item that may change over time. Updates are governed by an integer parameter~$\lambda$, that we refer to as the \textit{eviction threshold}: once an item is elected in a bucket, it remains so as long as its counter (scaled by $\lambda$) dominates the aggregate count of other items hashing to the same bucket. Upon an eviction from a bucket, the associated counters are reset, and the light block is updated to reflect the eviction. The estimated count of an item is obtained by adding its heavy-block counter (it is equal to $0$ if the item is not tracked there) to its light-block sketch-based estimate.


\begin{figure}
    \centering
    \includegraphics[width=0.4\linewidth, keepaspectratio]{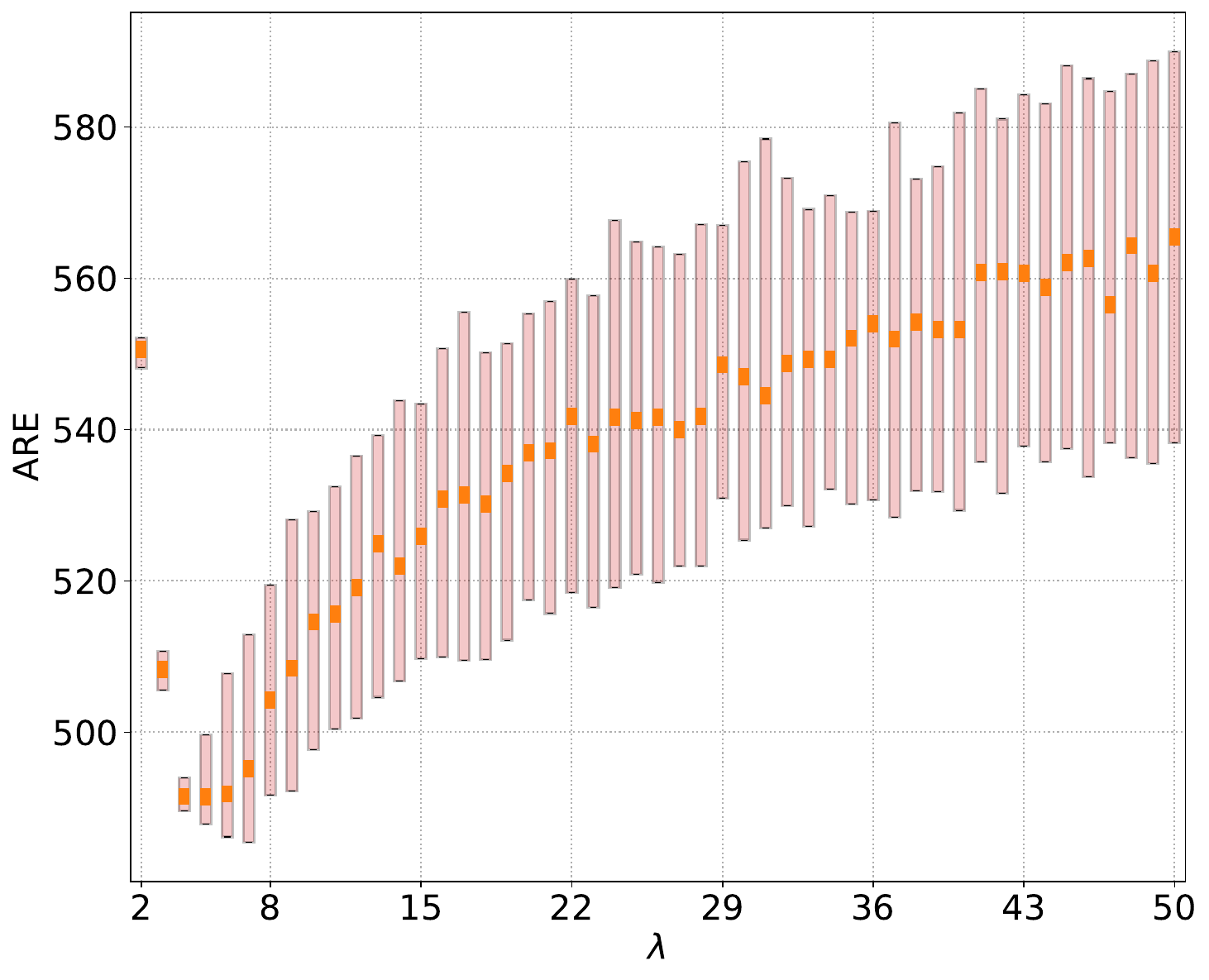}
   \caption{Average Relative Error (ARE) of \texttt{Elastic-Sketch} as a function of the eviction threshold $\lambda$, shown as box plots (boxes span Q1--Q3 and the center line indicates the median) over 100 runs, where each run is generated from an independent Zipf stream with skew parameter $\alpha=1.2$; $m_1=50$, $n_{\cI}=2\times 10^{5}$ items, stream length $\tau=5\times 10^{5}$, \texttt{CM} width $200$.}
\label{fig:are_n_runs100_m150_nItems20000_T500000_widthCMS_200_a12}
\end{figure}

$\ElaSk$'s estimation accuracy is highly sensitive to the characteristics of the input stream, including both the item frequency distribution and the arrival order, as well as the choice of eviction threshold~$\lambda$. For example, when~$\lambda\to\infty$, no evictions occur in the heavy block, causing the first item in each bucket to remain permanently elected---making performance extremely sensitive to arrival order. Even for moderate values of~$\lambda$, Figure~\ref{fig:are_n_runs100_m150_nItems20000_T500000_widthCMS_200_a12} demonstrates substantial variability in counting error across different realizations of streams with identical frequency distributions, highlighting the algorithm's sensitivity to arrival patterns.

This variability presents a fundamental challenge for practical deployment. The algorithm must be configured (particularly the choice of $\lambda$) before observing the actual stream, yet the optimal configuration depends heavily on unknown stream properties. 
However, real-world workloads often exhibit predictable statistical regularities, such as Zipf-like frequency distributions commonly observed in web traffic~\cite{breslau1999web,cao1997web}.
This suggests that, rather than optimizing for unknown worst-case scenarios, we should leverage these statistical patterns to guide algorithm configuration.

We therefore adopt a stationary random stream model where requests $\bm{R}=(R_t)_{t=1}^\tau$ are i.i.d. draws from a probability vector $\bm{p}=(p_i)_{i\in\mathcal I}$ over an item set $\mathcal{I}$ of size $n_{\mathcal{I}}$. This model captures the essential statistical properties of real workloads while remaining analytically tractable, enabling us to derive optimal parameter choices based on expected performance rather than pathological edge cases. Under this framework, we address the following questions:
\begin{center}
\begin{minipage}{0.9\linewidth}
\itshape
\begin{enumerate}
    \item What is the memory-accuracy trade-off of $\ElaSk$ as a function of the vector~$\bm{p}$? 
    \item How should one choose the memory split between the heavy and CM blocks (via $m_1$) and set the eviction threshold~$\lambda$ to minimize the expected counting error?

\end{enumerate}
\end{minipage}
\end{center}

\subsection{Contributions}

Our paper answers the above questions via three main contributions. First, we derive closed form expressions for the limiting distribution of the counters and the corresponding expected counting error. This permits efficient grid-based search for the optimal parameters $(\lambda,m_1)$ minimizing the expected counting error. Second, we characterize the structure of the optimal eviction threshold for a fixed $\beta$, substantially reducing the search space and showing how this threshold depends on the arrivals distribution~$\bm{p}$. Third, we validate our results on finite streams drawn from a Zipf distribution. We detail each contribution and summarize the techniques employed.







\paragraph{\textbf{Limiting behavior and near-optimal configuration}} Under the stationary random stream assumption, Theorem~\ref{thm:counters_limit} derives closed-form expressions of the counters limiting distribution, as the stream length tends to infinity. It shows that the asymptotic behavior in the heavy block is bucket-dependent and falls into two regimes: (i) the elected item switches infinitely often, or (ii) some item becomes permanently elected, with the election probability admitting a closed-form expression. 
Corollary~\ref{cor:limting_error_expectation} then yields closed-form expressions of the expected counting error. It shows that the expected counting error decreases with the aggregate probabilities of permanently elected items. For a fixed realization of the hash function~$\beta$, this expected error can be computed in~$\cO(n_{\cI})$ time. This enables efficient comparisons across $\ElaSk$'s configurations and near-optimal tuning of $\lambda$ and $m_1$ via grid search.

The proof of Theorem~\ref{thm:counters_limit} captures the evolution of each bucket via a discrete-time Markov chain that decomposes into asymmetric random-walk branches indexed by the currently elected item such that election switches correspond to transitions between branches. Recurrence versus transience of this chain yields regimes~(i) and~(ii) and provides explicit conditions for each.

\paragraph{\textbf{Characterization of the optimal eviction threshold}} Theorem~\ref{thm:lambda_star} shows that, for a fixed hash function~$\beta$, the optimal eviction threshold~$\lambda^{*}(\beta)$ that minimizes the expected counting error belongs to a finite candidate set of size at most~$m_1$. It also shows that a near-optimal threshold that minimizes the sample average counting error over $n_{\text{samp}}$ realizations of~$\beta$ can be chosen from at most~$m_1 \, n_{\text{samp}}$ candidate values. Finally, Theorem~\ref{thm:hp_lambda} provides probabilistic upper bounds on $\lambda^{*}(\beta)$ showing that, in most cases, the relevant candidates lie in a much smaller range than the worst-case bound~$m_1 \, n_{\text{samp}}$. Together, these results substantially reduce the cost of the grid search for tuning~$(\lambda, m_1)$.

The proof of Theorem~\ref{thm:lambda_star} shows that, for a fixed $\beta$, the expected counting error as a function of $\lambda$ is a piecewise continuous function and decreases on each piece. It then pinpoints the discontinuity values where the counting error can increase. The proof of Theorem~\ref{thm:hp_lambda} shows that the largest possible value of $\lambda^{*}(\beta)$ occurs when $\bm{p}$ is the uniform distribution. It also shows that~$\lambda^{*}(\beta)$ in this case reduces to the maximum load in a \textit{balls-and-bins} process, so classical results yield sharp high probability bounds.

\paragraph{\textbf{Finite-time validation.}} For a fixed hash function $\beta$, we validate our asymptotic results on Zipf streams of length~$\tau = 5\times 10^{5}$ with $n_{\cI}=10^{4}$ items and skew parameter~$\alpha\in \{ 0.8, 1.0, 1.2\}$. Figure~\ref{fig:g_lambda_zipf_alpha} shows that the theoretical limits closely match the corresponding simulation estimates at finite~$\tau$. Moreover, the candidate set for $\lambda^{*}(\beta)$ identified in Theorem~\ref{thm:lambda_star} is small in practice (less than $15$ values) despite the worst-case upper bound of~$m_1=200$. In the considered experimental setting, the asymptotically optimal eviction threshold $\lambda^{*}(\beta)$ is also optimal for the finite-time estimates. Moreover, the candidates set for the near-optimal eviction threshold, identified in Theorem~\ref{thm:lambda_star}, has less than $27$ elements, far below the worst-case upper bound $m_1 n_{\text{samp}}=2\times 10^{5}$ with $n_{\text{samp}}=10^3$.

\subsection{Related work}


 $\ElaSk$'s original paper~\cite{ElasticSketch2018} derives probabilistic bounds on the counting error that hold for a fixed stream. However, these bounds do not account for either the randomness of the hash function $\beta$ nor the arrivals. As a result, it is unclear how $ (\lambda,m_1)$ can be configured given only the items' frequency profiles. In contrast, our analysis provides an average-case performance metric in terms of the frequency profile, which can be directly used to tune $(\lambda,m_1)$. The authors in \cite{CormodeZipfErrorCMS2005} characterize the memory accuracy trade-off of $\CM$ under a Zipf frequency profile with skew $\alpha$. Specifically, they show that achieving~$\epsilon$ error in count estimation, with \texttt{CM}, requires~$\mathcal{O}(\epsilon^{-\min (1,1/\alpha)})$ memory. In $\ElaSk$, existing $\CM$ trade-offs apply to the $\CM$ block only after replacing the original stream by the \textit{filtered stream} obtained by removing updates absorbed by the heavy block. Hence, one needs to quantify the filtered-stream length (or, equivalently, the fraction of updates absorbed). This is precisely the hard part: unlike $\CM$, the heavy block is order dependent, so for a fixed frequency profile the absorbed mass can vary substantially across arrival orderings.

 Our work contributes to a line of research that derives tractable performance characterizations for streaming algorithms, with the additional goal of enabling principled configuration in practice. For instance, \cite{RecyclingBloomFilter_POMACS_24} analyzes recycling Bloom filters,~\cite{Mazziane_Marfoq_CMSCU25} studies $\CM$ with conservative updates, and~\cite{DISCO_dyanmic_config_d_BenBasat_ICDE24} proposes methods to optimize and adjust on the fly the number of hash functions in sketching algorithms.
 

 
 Finally, $\ElaSk$ was motivated by the fact that classical sketches struggle to adapt to line-rate processing on modern links. Moreover, a single sketch is often designed for a narrow family of queries, whereas network monitoring typically requires supporting a range of tasks, including per-item counts, frequency moments~\cite{braverman2014catchL2Sliding}, distinct counts~\cite{fusy2007estimatingCardinality}, entropy estimation~\cite{lall2006entropySigmetrics}, and change detection in traffic patterns. This has spurred a line of work---including $\ElaSk$---that augments or redesigns sketching primitives to reduce per-packet processing cost and broaden functionality. Representative examples include \texttt{UnivMon}~\cite{univmon16}, which builds a universal monitoring approach capable of answering multiple streaming queries; \texttt{SketchVisor}~\cite{sketch_visor_SIGCOMM17}, which introduces a fast path activated under high load; and \texttt{NitroSketch}~\cite{liu2019nitrosketch}, which lowers update cost by probabilistically updating only a subset of hashed counters rather than all of them.


\subsection{Notation}
 Vectors and matrices are written in bold, stream-induced random variables in uppercase, sets in calligraphic font. For any integer $k\ge 1$, $[k]\triangleq \{1,2,\ldots,k\}$, $\mathbb{N}$ designates the set of non-negative integers, $\mathds{1}(\cdot)$ denotes the indicator function, and $\setminus$ denotes the set difference operator. Probabilities and expectations with respect to the probability space governing the infinite stream $\bm{R}$ (when $\tau \to \infty $) are written as $\Proba{\cdot}$ and $\E{\cdot}$, respectively. When we also account for the randomness of the hash functions $(h_{\ell})_{\ell=1}^{d}$ and/or $\beta$, we write $\ProbaC{\bm{h}}{\cdot}$, $\ProbaC{\beta}{\cdot}$, and $\ProbaC{\bm{h},\beta}{\cdot}$, and analogously for expectations. The abbreviations a.s., w.h.p., and i.i.d. denote almost surely, with high probability and independent and identically distributed, respectively.  

\subsection{Paper Outline} 

The rest of the paper is organized as follows: \S~\ref{sec:problem} describes the \texttt{Elastic-Sketch} algorithm, introduces notation, and the adopted assumptions. \S~\ref{s:mains_result} presents the main results, \S~\ref{sec:simulations} validates them through numerical simulations, and \S~\ref{ss:Proof_counters_limit} provides the proof of Theorem~\ref{thm:counters_limit}. \S~\ref{s:Conclusion} concludes the paper. The appendix contains additional proofs and technical details.


\section{Problem Formulation} 
\label{sec:problem}

\begin{algorithm}
\algrenewcommand\algorithmicrequire{\textbf{Input:}}
\algrenewcommand\algorithmicensure{\textbf{Output:}}
\caption{\texttt{Elastic-Sketch}}
\label{Alg:Elastic_Sketch}
\begin{algorithmic}[1]
\Require Stream $\bm{R}=(R_t)_{t=1}^T$, hash functions $\beta, (h_{\ell})_{\ell\in[d]}$, eviction threshold $\lambda$, 
\Ensure Estimated count $\hat{\textbf{N}}$
\State Initialize $\bm{V}^+(0) \gets \bm{0}$, $\bm{V}^-(0) \gets \bm{0}$, $\bm{S}(0) \gets \bm{-1}$, $\bm{Y}(0) \gets 0$
\For{$t = 1$ to $T$}
    \State $(\bm{S}(t), \bm{V}^{+}(t), \bm{V}^{-}(t))\gets (\bm{S}(t-1), \bm{V}^{+}(t-1), \bm{V}^{-}(t-1))$
    \State $\bm{Y}(t) \gets \bm{Y}(t-1)$
    \State $b \gets \beta(R_t)$
    \If{$S_{b}(t-1)=-1$ or $S_{b}(t-1) = R_t$ }
        \State $S_{b}(t)\gets R_t$
        \State $V^+_{b}(t) \gets V^+_{b}(t-1) + 1$
        \Comment{Track in the heavy block}
    \Else
        \If{{$  \lambda V^{+}_{b}(t-1) - V_{b}^{-}(t-1)> 0$}} \label{line:updates_S}
            \State $S_{b}(t) \gets S_{b}(t-1)$
            \State $V^-_{b}(t) \gets V^-_{b}(t-1) + 1$
            \State $Y_{\ell,h_{\ell}(R_t)}(t)\gets Y_{\ell,h_{\ell}(R_t)}(t-1) +1$, $\forall l\in [d]$ \Comment{Track in \texttt{CM}}
        \Else  \label{line:Eviction_Elsk}
            \State $Y_{\ell,h_{\ell}(R_t)}(t)\gets Y_{\ell,h_{\ell}(R_t)}(t-1) +V^+_{b}(t-1)$, $\forall l\in [d]$ \Comment{Insert into \texttt{CM}}
            \State $S_{b}(t) \gets R_t$ \Comment{Replace bucket item}
            \State $V^+_{b}(t) \gets 1$
            \State $V^-_{b}(t) \gets 0$
        \EndIf
    \EndIf
    \State $\hat N_i(t) \triangleq \mathds{1}\left( i\in \bm{S}(t) \right) V^{+}_{\beta(i)} (t) + \min_{\ell\in [d]} Y_{\ell,h_{\ell}(i)} (t), \, \forall i\in \cI$ 
\EndFor
\end{algorithmic}
\end{algorithm}

Consider a data stream $\bm{R}=(R_1, \ldots, R_{\tau})$, of size~$\tau$, from a finite set~$\cI$ of size~$n_{\cI}$. For each item $i\in\cI$ and time $t$, let $N_i(t)$ be the number of occurrences of $i$ up to step $t$, and let $\bm{N}(t)\triangleq (N_i(t))_{i\in\cI}$ be the corresponding count vector. We assume that each arrival is an i.i.d. draw from a probability distribution~$\bm{p}=(p_i)_{i\in\cI}$. Formally, 
\begin{assumption}[Random Stationary Stream]\label{assum:stationary_stream} 
    The random variables $(R_t)_{t=1}^{\tau}$ are i.i.d. such that the probability that $R_1=i$, for any $i\in \cI$, is equal to $p_i$, i.e., $\Proba{R_1 = i} = p_i$, and $p_i>0$.  
\end{assumption}
As mentioned in the introduction, \texttt{Elastic-Sketch}'s performance depends strongly on the stream, making worst-case analysis of limited practical relevance and motivating an average-case study under a \textit{distribution} over streams.

 \texttt{Elastic-Sketch}, among other streaming algorithms, aims to answer queries related to the number of appearances of items in the stream, such as the detection of $\phi$-heavy hitters and estimation of $\bm{N}(t)$'s norms~\cite{ElasticSketch2018}. It combines a \textit{heavy} block, for monitoring popular items, with a Count-Min Sketch (\texttt{CM}) block. We describe the dynamics of each block, as specified in Alg.~\ref{Alg:Elastic_Sketch}.  

\medskip 

\noindent \textbf{Heavy block.} This block uses a parameter $\lambda$, called the \textit{eviction threshold}, and maintains a set~$\mathcal{B}$ of~$m_1$ buckets indexed by a hash function $\beta: \mathcal{I} \mapsto \mathcal{B}$---in practice the hash function produces an integer in~$[m_1]$, such that each integer identifies a bucket $b\in \mathcal{B}$. 
Each bucket~$b$ maintains three components: the currently \textit{elected} item~$S_b$, a counter~$V_b^{+}$ that tracks~$S_b$, and a counter~$V_b^{-}$ used in the eviction rule for~$S_b$. Define $\cI_{b}$ as the set of items hashing to $b$, i.e., $\cI_{b} \triangleq \{ i \in \mathcal{I} : \beta(i) = b \}$ and~$n_b$ as its size ($n_b\triangleq |\cI_b|$).

The elected item for each bucket~$b$ changes dynamically. An epoch for~$b$ is the interval from the moment an item is elected as~$S_b$ until its eviction. Throughout the epoch, upon each arrival mapped to~$b$ ($\beta(R_t)=b$ for some $t$), two counters are updated: $V_b^{+}$ (incremented on occurrences of $S_b$) and $V_b^{-}$ (incremented on occurrences from items in $\cI_{b}\setminus \{S_b\}$). The elected item changes when an item in $\cI_{b}\setminus \{S_b\}$ arrives while $\lambda V_b^{+}$ is equal to $V_b^{-}$. Formally, at $t=0$, $S_b(0)=-1$, modeling the fact that no item is monitored, $V_{b}^{+}(0)=0$, and $V_{b}^{-}(0)=0$. At each step $t\geq 1$, the algorithm computes $b(t)=\beta(R_t)$. Only the bucket~$b(t)$ is updated at step $t$, i.e., 
    \begin{align}
            (S_{b}(t),V_{b}^{+}(t),V_{b}^{-}(t)) =  (S_{b}(t-1),V_{b}^{+}(t-1),V_{b}^{-}(t-1)) , \; \forall b \in \mathcal{B}\setminus \{b(t)\}. 
    \end{align}



\noindent We distinguish three cases. 
\begin{enumerate}
        \item $S_{b(t)}(t-1)\in \{ -1 , R_t\}$, i.e., there is no currently monitored item in the bucket $b(t)$, or $R_t$ is itself the monitored item. \label{case1:ES}

        \item $S_{b(t)}(t-1)\notin \{ -1 , R_t\}$, i.e., there is an item monitored in bucket $b(t)$, but it is not $R_t$, and $\lambda V_{b(t)}^{+}(t-1) > V_{b(t)}^{-}(t-1)$. \label{case2:ES}

        \item $S_{b(t)}(t-1)\notin \{ -1 , R_t\}$, and $\lambda V_{b(t)}^{+}(t-1) = V_{b(t)}^{-}(t-1)$. \label{case3:ES}
\end{enumerate}

\noindent The update of bucket $b(t)$ depends on these three cases. In case~\ref{case1:ES}, $S_{b(t)}$ starts/keeps monitoring $R_t$, and $V_{b(t)}^{+}$ is incremented by $1$. In case~\ref{case2:ES}, the bucket $b(t)$ keeps monitoring $S_{b(t)}(t-1)$, and in this case it is $V_{b(t)}^{-}$ that is incremented by $1$ and not $V_{b(t)}^{+}$. In case~\ref{case3:ES}, $S_{b(t)}$ changes value to $R_t$, and the counts in $V_{b(t)}^{+}$ and $V_{b(t)}^{-}$ are reset to~$1$ and~$0$, respectively. Note that by construction of the algorithm, it is always true that $\lambda V_b^{+}(t)- V_{b}^{-}(t)\geq 0$, and thus all cases are covered. Formally, 
\begin{align}\label{e:Update_Block1}
(S_{b(t)}(t),V_{b(t)}^{+}(t),V_{b(t)}^{-}(t))=
    \begin{cases}
        (R_t, V_{b(t)}^{+}(t-1) + 1, V_{b(t)}^{-}(t-1) ),  & \text{ if case \ref{case1:ES},}  \\ 
        (S_{b(t)}(t-1), V_{b(t)}^{+}(t-1), V_{b(t)}^{-}(t-1) +1 ), & \text{ if case \ref{case2:ES},} \\ 
        (R_t, 1, 0), & \text{ if case \ref{case3:ES}.}
    \end{cases}
\end{align}

It will be useful later to define, for each bucket~$b$, a vector notation for bucket indexed processes $S_b$, $V_b^{+}$, and $V_b^{-}$, e.g.,  $\bm{S}(t)\triangleq(S_b(t))_{b\in \cB}$, 
$\mathcal{E}_{\beta}(t)$ as the set of elected items at step $t$, i.e., $\mathcal{E}_{\beta}(t)\triangleq \{S_b(t): b\in \cB \}$,
and $\mu_b \triangleq \sum_{i\in \cI_{b}} p_i$. Further define for each item $i$, 
    \begin{align}
     \lambda_{i}(\beta) \triangleq \frac{\mu_{\beta(i)}}{p_i}  - 1,  
    \end{align}
and $\lambda_{b}^{(k)}(\beta)$ as the $k$-th smallest value of the set $\{\lambda_i(\beta): i\in\cI_b\}$, for any $b\in \cB$, e.g., \\ 
$\lambda_b^{(1)}(\beta)=~\min_{i\in \cI_b} \lambda_i(\beta)$. Similarly, we define $p_b^{(k)}$ and $N_{b}^{(k)}(t)$ as the $k$-th largest values of the sets $\{p_i: i\in\cI_b\}$  and $\{N_i(t): i\in\cI_b\}$, respectively.


\medskip

\noindent \textbf{\texttt{CM} block.} This block uses Count-Min Sketch (\texttt{CM})~\cite{cormode2005improved} to summarize counts of items not currently monitored in the heavy block, i.e., items in~$\cI \setminus \{ S_b(t): b\in \cB\}$. We denote the value of the counters matrix of \texttt{CM} within \texttt{Elastic-Sketch} as~$\bm{Y}(t) = (Y_{\ell,c}(t))_{(r,c)\in [d]\times[m_2]}$ where $d$ and $m_2$ are the number of rows and columns.

In the absence of the heavy block ($m_1=0$), \texttt{Elastic-Sketch} reduces to \texttt{CM}. In this case, we denote the counters matrix of the \texttt{CM} block as  $\bm{Y}^{\texttt{CM}}(t)$. \texttt{CM} uses $d$ hash functions, $h_{\ell}: \cI \mapsto [m_2]$ for $l\in [d]$, to map items to counters. Initially, $Y_{\ell,c}^{\texttt{CM}}(0)=0$, and at any step $t\geq 1$, all counters associated to $R_t$, $(l,h_{\ell}(R_t))_{\ell=1}^{d}$, are incremented,
\begin{align}
        Y_{\ell,c}^{\texttt{CM}}(t) = Y_{\ell,c}^{\texttt{CM}}(t-1) + \mathds{1}\left( h_{\ell}(R_t) = c\right). 
\end{align}
On the other hand, when $m_1>0$, the update of~$\bm{Y}$ depends on the three cases distinguished for the heavy block. The matrix~$\bm{Y}$ is not updated on hits to the currently elected item at bucket~$b(t)$ (case~\ref{case1:ES}), but it is updated on arrivals of items mapped to~$b(t)$ that are not~$S_{b(t)}(t-1)$. In case~\ref{case2:ES}, $R_t$ is added to the sketch with value~$1$. Upon an eviction of the elected item at bucket~$b(t)$ (case~\ref{case3:ES}), the epoch ends: $V_b^{+}(t-1)$ holds the evicted item’s total during the epoch, so this amount is added to the sketch. Formally, for any counter $(l,c)$,
\begin{align}\label{e:Update_Block2}
        Y_{\ell,c}(t)=
    \begin{cases}
        Y_{\ell,c}(t-1),  & \text{ if case \ref{case1:ES},}  \\ 
        Y_{\ell,c}(t-1) + \mathds{1}\left( h_{\ell}(R_t) = c\right), & \text{ if case \ref{case2:ES},} \\ 
        Y_{\ell,c}(t-1) + \mathds{1}\left( h_{\ell}(R_t) = c\right) \cdot V_{b(t)}^{+}(t-1), & \text{ if case \ref{case3:ES}.}
    \end{cases}
\end{align}

\medskip 

\noindent \textbf{Count estimation.} In \texttt{CM}, at any step $t$, the estimate for item~$i$ is the minimum of its~$d$ counters, namely $\min_{\ell\in [d]} Y^{\texttt{CM}}_{\ell,h_{\ell}(i)}(t)$, since each cell $(r,h_r(i))$ overestimates $N_i(t)$ and the minimum is the tightest. In Elastic Sketch, the same rule applies to items not currently monitored in the heavy block. However, if an item $i$ is currently elected at some $b\in \cB$, arrivals of $i$ during the current epoch do not update the sketch; instead, they accumulate in $V_b^{+}(t)$. To preserve overestimation and account for this, \texttt{Elastic-Sketch} adds $V_b^{+}(t)$ to the sketch-based estimate. In summary, the estimation count for item~$i$ in \texttt{Elastic-Sketch} at step~$t$, denoted~$\hat N_i(t)$, is given by, 
    \begin{align}
            \hat N_i(t) \triangleq \mathds{1}\left( i\in \mathcal{E}_{\beta}(t) \right) V^{+}_{\beta(i)} (t) + \min_{\ell\in [d]} Y_{\ell,h_{\ell}(i)} (t).
    \end{align}

\texttt{Elastic-Sketch} is a randomized algorithm, due to the random choice of the hash functions~$\beta$ and~$(h_{\ell})_{\ell=1}^{d}$. 

\begin{assumption}[Ideal Hash Functions Model]\label{assum:IdealHash}
        For any $\ell\in [d]$, the sets~$(\beta(i))_{i\in \mathcal{I}}$ and $\left( h_{\ell}(i)\right)_{i\in \mathcal{I}}$ are i.i.d. uniform random variables in $[m_1]$ and $[m_2]$ respectively.
\end{assumption}
While constructing hash functions that satisfy Assumption~\ref{assum:IdealHash} can be costly, such idealized models have been shown to accurately predict the performance of algorithms using practical hash functions with weaker independence guarantees~\cite{mitzenmacher2008simple}, and are widely adopted in the analysis of hash-based data structures and algorithms~\cite{BroderBloomFilter2003,mitzenmacher2017probability}.


\medskip 

\noindent \textbf{Performance metric.} Let~$\Err_i(t)$ be the difference between the estimation and the true count of item~$i$ at step~$t$, i.e., $\Err_{i}(t) = \hat N_i(t) - N_{i}(t)$. If an item $i_0$ is absent up to time $t$ ($R_s\neq i$ for all $s\leq t$), then $\Err_{i_0}(t)=\min_{\ell\in [d]} Y_{\ell,h_{\ell}(i_0)}$. Under Assumption~\ref{assum:IdealHash}, $\{h_{\ell}(i_0)\}_{\ell\in [d]}$ are independent of $\bm{Y}(t)$, whose randomness is determined by~$\{h_{\ell}(R_s)\}_{s\leq t, \; l\in [d]}$ and $\beta$. Hence $\Err_{i_0}(t)$ is identically distributed as the minimum of~$d$ row-wise uniformly sampled counters. We define the counting error $\Err_{(0)}(t)$ as follows, 
\begin{align}\label{e:def_UE}
            \Err_{(0)}(t) \triangleq  \min_{\ell\in [d]}   Y_{\ell, X_{\ell}}(t),
\end{align}
where $(X_{\ell})_{\ell\in [d]}$ are i.i.d. uniform random variables over~$[w]$. It is clear then that, for any item~$i_0$ absent from the stream, $\Err_{i_0}(t)$ is identically distributed to~$\Err_{(0)}(t)$. Moreover, taking $X_{\ell}=h_{\ell}(i)$ for any item $i$, yields that, $\Err_{(0)}(t) - \Err_i(t) \geq N_{i}(t) -V_{\beta(i)}^{+} \geq 0$.
Thus $\Err_{(0)}$ \textit{stochastically dominates}~$\Err_i$ for any~$i$, i.e., 
    \begin{align}\label{e:SD_e0_ei}
            \Proba{\Err_{i}(t)\geq x}~\leq~\Proba{\Err_{(0)}(t)\geq x}, \; \forall x, t. 
    \end{align}
The quantity~$\Err_{(0)}$ provides an upper bound on the estimation error for any item, and thus it can be used to bound standard performance metrics, such as the \textit{average relative error}~\cite{ElasticSketch2019} and the expected false positive rate in the detection of~$\phi$-heavy-hitters.

The next section derives the limiting distribution of~$\Err_{(0)}(t)/t$ under the random stationary stream model, thereby quantifying the memory-accuracy trade-off of \texttt{Elastic-Sketch} and yielding practical guidelines for tuning $\lambda$ and $m_1$.

\section{Main Results} 
\label{s:mains_result}

This section is organized as follows. Section~\ref{ss:Limiting_distribution} derives the limiting distribution of the counters and the resulting asymptotic counting error. These expressions yield efficient numerical procedures for near-optimal tuning of $(\lambda,m_1)$ via a grid-search. Section~\ref{ss:optimal_eviction_threshold} further characterizes the optimal eviction threshold that minimizes the expected limiting counting error. This helps reduce the search space and offers insights about the optimal threshold for the arrival distribution $\bm{p}$.





\subsection{Limiting distribution of the counters} 
\label{ss:Limiting_distribution}

As highlighted in Fig.~\ref{fig:are_n_runs100_m150_nItems20000_T500000_widthCMS_200_a12}, $\ElaSk$'s strong sensitivity to the stream ordering is primarily due to the heavy block. Indeed, when the heavy block is absent ($m_1=0$), $\ElaSk$ reduces to $\CM$, and under the random stationary stream model (Assumption~\ref{assum:stationary_stream}), the \textit{Strong Law of Large Numbers} yields $\frac{1}{t} Y_{\ell,c}^{\texttt{CM}}(t) \xrightarrow{a.s.}  \sum_{i\in \mathcal{I}} \mathds{1}(h_{\ell}(i)=c) p_i$, when $t$ tends to infinity. Thus, all stream realizations regardless of ordering, lead to the same asymptotic counter values. 

Unfortunately, previous analysis of $\ElaSk$ fails to capture its strong dependency on the stream. Indeed, \cite[Thm. 4]{ElasticSketch2019} derives a probabilistic upper bound on $\Err_i$ for a fixed stream, over the randomness of the hash functions~$(h_{\ell})_{\ell=1}^{d}$. The bound is linear in~$t-V_{\cB}^{+}(t)$, where $V_{\cB}^{+}(t) \triangleq~\sum_{b\in\cB} V_b^{+}(t)$.
However, this offers limited insight into \textit{average} performance because, even among streams with the same final count vector, $V_{\cB}^{+}(t)$ can vary from~$m_1$ up to~$\sum_{b\in \cB}N^{(1)}_b (t)$. The minimum occurs when the arrivals of popular items are dispersed, triggering frequent evictions of items from the heavy block, and resets of~$V_b^{+}$ (see \eqref{e:Update_Block1}). The maximum occurs when they arrive early in bursts, so $V_b^{+}$ accumulates with time and there are no evictions.



To gain insight into the average performance of $\ElaSk$, we study the asymptotic distribution of $S_b(t)$, $V_{b}^{+}(t)/t$, and $Y_{\ell,c}(t)/t$, under the random stationary stream assumption. In order to describe our result we introduce a few notations and definitions. Define $\cB_{\beta}^{+}(\lambda)$ as the set of buckets $b$ where either $\lambda_{i}(\beta)$ is strictly smaller than~$\lambda$ for any item~$i$ with~$\beta(i)=b$, or for which there is a single item hashing to it, i.e., $n_b=1$. Likewise, let $\cB_{\beta}^{0}(\lambda)$ be the set of buckets $b$ with $n_b=0$, and $\cB_{\beta}^{-}(\lambda)$ be the remaining buckets. Formally,  
    \begin{align} \nonumber 
    &\cB_\beta^0(\lambda) \triangleq \{b\in\cB: \cI_b= \emptyset \}, \; 
     \cB_{\beta}^{+}(\lambda) \triangleq \left\{ b\in \cB: \; \lambda_{b}^{(1)}(\beta)< \lambda, \text{ or } n_b = 1 \right\}, \\
     & \quad \quad \quad \quad \quad \cB_{\beta}^{-}(\lambda)\triangleq \cB \setminus \left(\cB^{+}_{\beta}(\lambda) \cup \cB_{\beta}^{0}(\lambda)\right).
          \label{eq:bucket_sets}
    \end{align}
Let $\phi: [0,1]\times [1,+\infty) \mapsto \mathbb{R}^{+}$ be defined as $\phi(x,\lambda) \triangleq \frac{x^{\lambda+1} -1}{x-1}$ and let \begin{align}\label{e:def_r_root_function}
      r(\lambda, z) \triangleq
      \begin{cases}
              1, &\text{ if }  \lambda \leq z-1,  \\ 
              \text{root of } \phi(\cdot, \lambda) - z \text{ in } [0,1],  &  \text{ otherwise,}
      \end{cases}.
\end{align}
Define the functions~$w: [1,+\infty)\times [1,+\infty] \mapsto \mathbb{R}^{+}$ and $w_{i,\beta}: [1,+\infty) \mapsto \mathbb{R}^{+}$ as, 
\begin{align}\label{e:def_absorption_weights}
      w(\lambda, z) \triangleq \left( 1- (r(\lambda,z))^{\lambda} \right), \quad \; \text{ and } \; \quad w_{i,\beta}(\lambda) \triangleq p_i w\left(\lambda, \frac{\mu_{\beta(i)}}{p_i}\right). 
\end{align}

\noindent Observe that $w_{i,\beta}(\lambda)>0$ for any~$\lambda> -1 + \nicefrac{\mu_{\beta(i)}}{p_i}  = \lambda_{i}(\beta)$.
 
Theorem~\ref{thm:counters_limit} asserts that for any bucket~$b\in \cB_{\beta}^{+}(\lambda)$, an item $i\in \cI_b$ eventually becomes permanently elected, whereas for any bucket $b\in \cB_{\beta}^{-}(\lambda)$, $S_b$ switches states infinitely often. Define
\begin{equation}
\label{e:Sbinfty}
    S_b^\infty \triangleq 
    \begin{cases}
        \lim_{t\to \infty} S_b(t) & \text{ for } b\in \cB_{\beta}^{+}(\lambda), \\
        -1 & \text{ for } b\in \cB_{\beta}^{-}(\lambda)\cup \cB_{\beta}^{0}(\lambda).
    \end{cases}
\end{equation}
Additionally, let $p_{-1} \triangleq 0$.

The theorem also establishes that for any item~$i\in \cI$ the probability that $S_b^{\infty}=i$, is equal to~$a_{i,\beta}(\lambda)$, where $a_{i,\beta} : (0,+\infty)\mapsto \mathbb{R}$ is a piecewise function defined as,
\begin{align}\label{e:def_absorption_proba}
    a_{i, \beta}(\lambda) \triangleq
    \begin{cases}
     1, & \text{ if } n_{\beta(i)}=1, \\
     0, & \text{ if } \lambda \in \left( 0, \lambda_{i}(\beta)\right] \text{ and } n_{\beta(i)}\geq 2,  \\
    \frac{w_{i,\beta}(\lambda)}{\sum_{j: \;\beta(j)=\beta(i)} w_{j,\beta}(\lambda)}, & \text{ else.}
    \end{cases}
\end{align}  



\begin{theoremSp}[Counters limiting distribution]
\label{thm:counters_limit}
Under Assumption~\ref{assum:stationary_stream}, 
\begin{enumerate}
\item For any bucket $b\in \cB^{-}_{\beta}(\lambda)$, $(S_b(t))_t$ switches state infinitely often, and for any bucket $b\in \cB^{+}_{\beta}(\lambda)$, $S_b(t)$ converges a.s. in $t$, such that for any item $i\in \cI_b$, 
\begin{align}\label{e:distribution_S_infinity}
  \Proba{S_{b}^{\infty}=i}=  a_{i,\beta}(\lambda),
\end{align} 
\label{item:thm_limit_S}
where $a_{i,\beta}(\lambda)$ is defined in \eqref{e:def_absorption_proba}.
\item For any bucket $b\in \cB$, the counter $V_b^{+}$ satisfies, 
\begin{align}\label{e:limit_V_plus}
\frac{V_{b}^{+}(t)}{t}  \xrightarrow[t\to \infty]{a.s.}  \mathds{1}\left( b\in \cB_{\beta}^{+}(\lambda) \right) p_{S_{b}^{\infty}}.
\end{align} \label{item:thm_limit_V}
\item For any cell $(\ell,c)$ in the $\CM$ block, the counter $Y_{\ell,c}$ satisfies,
\begin{align}\label{e:limit_y} 
& \frac{Y_{\ell,c}(t)}{t} \xrightarrow[t\to \infty]{a.s.} \sum_{i\in \mathcal{I}} \mathds{1}\left(h_{\ell}(i) = c\right) p_i - \sum_{b\in \cB_{\beta}^{+}(\lambda)} \mathds{1}\left(h_{\ell}(S_b^{\infty}=c)\right) p_{S_b^{\infty}}.
\end{align} \label{item:thm_limit_Y}
\end{enumerate}
\end{theoremSp}

\begin{proof}[Sketch of the proof]

To prove item~\ref{item:thm_limit_S} of the theorem, we capture the evolution of each bucket as a discrete-time Markov chain $M_b(t)\triangleq(S_b(t), U_b(t))$ such that $U_b(t)\triangleq \lambda V_b^{+}(t)- V_b^{-}(t)$. The chain decomposes into asymmetric random-walk branches indexed by the currently elected item, and election switches correspond to transitions between branches. The long-run stability of the elected item is then characterized by the transience/recurrence of this chain: recurrence leads to infinitely many switches ($b\in \cB_{\beta}^{-}$), whereas absorption in a branch yields a permanently elected item ($b\in \cB_{\beta}^{+}$). In the latter case, computing the branch absorption probabilities in $M_b$ yields~\eqref{e:distribution_S_infinity}.

To prove items~\ref{item:thm_limit_V} and~\ref{item:thm_limit_Y} of the theorem, we first establish in Lemma~\ref{lem:Count_ElasticSketch_Finite} (Section~\ref{ss:Proof_counters_limit}) a finite-time characterization of the counter values in terms of the count vector~$\bm{N}(t)$ and the last eviction time in each bucket. We then combine this characterization with the asymptotic behavior of $(S_b(t))$ to derive the corresponding asymptotic results for the counters.

The detailed proof is presented in Section~\ref{ss:Proof_counters_limit}. 
\end{proof}

\begin{algorithm}
\algrenewcommand\algorithmicrequire{\textbf{Input:}}
\algrenewcommand\algorithmicensure{\textbf{Output:}}
\caption{Pseudo code for the computation of $g_{\beta}(\lambda)$ }
\label{alg:gbeta_computation}
\begin{algorithmic}[1]
\Require Items set $\cI$, probabilities vector $\bm{p}$, hash function $\beta$, eviction threshold $\lambda$
\Ensure $g_{\beta}(\lambda)$
\State Build bucket lists $(\cI_b)_{b\in\cB}$ by hashing each $i\in\cI$ \Comment{$\mathcal{O}(n_{\cI})$}
\State Compute $\mu_b\gets \sum_{i\in\cI_b} p_i$ for all $b\in\cB$ \Comment{$\mathcal{O}(n_{\cI})$}
\State Compute $\lambda_{b}^{(1)}(\beta)$ for all $b\in\cB$ \Comment{$\mathcal{O}(n_{\cI})$}
\State $g\gets 0$
\For{$ b\in \cB: \; (n_b\geq 1 \text{ and } \lambda_b^{(1)}(\beta) < \lambda )$} \Comment{$\mathcal{O}(n_b)$ per bucket}
        \State Compute $w_{i,\beta}(\lambda)$ for all $i\in\cI_b$ using \eqref{e:def_absorption_weights}--\eqref{e:def_r_root_function} 
        \State $s_{p}\gets \sum_{i\in\cI_b} w_{i,\beta}(\lambda) p_i $
        \State $s\gets \sum_{i\in\cI_b} w_{i,\beta}(\lambda)$  
        \State $g\gets g+ \nicefrac{s_p}{s}$ 
\EndFor
\State \Return $g$ \Comment{Time complexity $\mathcal{O}(n_{\cI})$} 
\end{algorithmic}
\end{algorithm}

When the $\CM$ block uses a single hash function, i.e., $d=1$, Corollary~\ref{cor:limting_error_expectation} 
characterizes the expected limiting counting error, $\aErr^{\infty} \triangleq~\lim_{t\to \infty}\min_{\ell\in [d]} Y_{\ell,X_{\ell}}(t)/t$ in terms of the function $g_{\beta}:[1,+\infty)\to\mathbb{R}^{+}$, defined as, 
\begin{align} \label{e:def_g_and_gb}
  g_{\beta}(\lambda) 
  \triangleq \sum_{b\in \cB^{+}_{\beta}(\lambda)} g_{b,\beta}(\lambda),
  \quad \text{ where } \quad
  g_{b,\beta}(\lambda) \triangleq \sum_{i\in \cI_{b}} a_{i,\beta}(\lambda) \,  p_i.
\end{align}

\begin{corollarySp}[Expected Limiting Counting Error]\label{cor:limting_error_expectation} Under Assumption~\ref{assum:stationary_stream} and when $d=1$, the following holds, 
    \begin{align}
    \Esub{\bm{h},\beta}{\aErr^{\infty}} = \frac{1}{m_2} \left( 1 - \Esub{\beta}{g_{\beta}(\lambda)} \right).
\end{align}
\end{corollarySp}


For any $d\geq 1$, the same right-hand side remains a valid upper bound, since $\CM$ estimate is a minimum over~$d$ rows and thus is upper-bounded by any single-row estimate. Importantly, the choice $d=1$ aligns with implementation considerations in~\cite[Sec. 4]{ElasticSketch2019}, where it is recommended to favor throughput over marginal accuracy gains. In the absence of the heavy block, the corresponding term equals~$1/m_2$. Hence, the quantity~$\left( 1 - \Esub{\beta}{g_{\beta}(\lambda)} \right)\in [0,1]$ quantifies the average reduction relative to the baseline~$1/m_2$.

Given an arrival distribution~$\bm{p}$, Corollary~\ref{cor:limting_error_expectation} provides an explicit asymptotic objective for evaluating and tuning~$\ElaSk$. It enables direct comparison of configurations jointly in the eviction threshold $\lambda$ and the memory split between the heavy and $\CM$ blocks, parametrized by $m_1$ and $m_2$. A near-optimal configuration for $\bm{p}$ can then be obtained numerically by grid searching over $(\lambda, m_1, m_2)$, and maximizing the Monte Carlo estimate, 
\begin{align}\label{e:approx_expectation_g_beta}
\widehat{\mathbb{E}}^{n_{\text{samp}}}_{\beta}\left[ g_{\beta}(\lambda) \right] \triangleq
    \frac{1}{n_{\text{samp}}} \sum_{k=1}^{n_{\text{samp}}} g_{\beta_k}(\lambda),
\end{align}
where the hash functions $\beta_k$ are obtained by varying the hash seed and $n_{\text{samp}}$ is the total number of seeds. As $n_{\text{samp}}$ grows, \eqref{e:approx_expectation_g_beta} converges to $\Esub{\beta}{g_{\beta}(\lambda)}$, yielding the configuration that minimizes the expected limiting counting error. Note that even under an ideal hashing model such as Assumption~\ref{assum:IdealHash}, computing $\Esub{\beta}{g_{\beta}(\lambda)}$ exactly is intractable because of the large space of hash functions and the non-linearity of the function $g_{\beta}$. A Monte Carlo approach as described above is more suitable since it estimates performance under the hash family used in practice, via different seeds, rather than under an idealized hashing model.

The brute-force tuning of~$\ElaSk$ over a grid of configurations~$(\lambda, m_1)$ has time complexity~$\cO(n_{\text{grid}} \, n_{\text{samp}} \, n_{\cI})$, where~$n_{\text{grid}}$ is the number of tested couples. This follows because Theorem~\ref{thm:counters_limit} provides closed form expressions for the election probabilities~$a_{i,\beta}(\lambda)$, yielding an $\mathcal{O}(n_{I})$ time computation of~$g_{\beta}(\lambda)$, as shown in Alg.~\ref{alg:gbeta_computation}. While under a fixed memory budget, the feasible values of~$m_1$ are bounded, $\lambda$ can in principle be arbitrarily large and may therefore dominate~$n_{\text{grid}}$. To address this, in the next section, we derive properties of the optimal eviction threshold for a given~$\beta$. These properties restrict the search range for $\lambda$.

\subsection{Optimal eviction threshold}
\label{ss:optimal_eviction_threshold}

For a fixed value of $m_1$, define~$\lambda^{*}$ as the smallest eviction threshold that maximizes the expectation of the function~$g_{\beta}(\lambda)$, and equivalently minimizes the expected limiting counting error when~$d=~1$, as shown in Corollary~\ref{cor:limting_error_expectation}. 
We settle instead for an approximation of~$\lambda^{*}$, denoted $\widehat{\lambda}^{*}$, and defined as, 
\begin{align}\label{e:lambdahatstar}
\widehat{\lambda}^{*} \triangleq \min  \argmax_{\lambda \in \mathbb{N}\setminus\{0\}}  \widehat{\mathbb{E}}^{n_{\text{samp}}}_{\beta}\left[g_{\beta}(\lambda)\right]. 
\end{align}
Similarly, we define~$\lambda^{*}(\beta)$ as the smallest optimal eviction threshold for a fixed realization of the hash function $\beta$, i.e.,  
    \begin{align}
            \lambda^{*}(\beta) \triangleq \min \argmax_{\lambda \in \mathbb{N}\setminus\{0\}}  g_{\beta}(\lambda).
    \end{align}
We further define the set $\Lambda(\beta)$ as follows, 
\begin{align}
        \Lambda(\beta) \triangleq \left\{ \floor{\lambda^{(1)}_b(\beta)} + 1 : \; b\in \cB \right\}.
\end{align}

Theorem~\ref{thm:lambda_star} narrows down the search for $\lambda^{*}(\beta)$ and $\widehat{\lambda}^{*}$ to at most $m_1$ and $m_1\, n_{\text{samp}}$ candidate values, respectively. The proof uses Lemma~\ref{lem:g_b_decreasing}. 

\begin{lemmaSp}\label{lem:g_b_decreasing}
   For any bucket $b\in \cB$ The function $g_{b,\beta}$ is decreasing over the interval $\left(\lambda_{b}^{(1)}(\beta),+\infty\right)$. 
\end{lemmaSp}
\begin{proof}
    The proof is presented in Appendix~\ref{app:proof_thm_lambda_star}. 
\end{proof}

\begin{theoremSp}[Candidate values of $\lambda^{*}(\beta)$ and $\widehat{\lambda}^{*}$] \label{thm:lambda_star} 
The following holds, 

    \begin{enumerate}
        \item For a fixed $\beta$, the optimal eviction threshold $\lambda^{*}(\beta)$ satisfies,
        \begin{align}\label{e:candidates_lambda_star}  
\lambda^{*}(\beta)\in \Lambda(\beta).
\end{align}    

    \item The near-optimal eviction threshold  $\widehat{\lambda}^{*}$ satisfies, 
    \begin{align}
            \widehat{\lambda}^{*} \in \bigcup_{k=1}^{n_{\text{samp}}} \Lambda(\beta_k).
    \end{align}
    \end{enumerate}
\end{theoremSp}
\begin{proof}
 For each bucket~$b$, from the definition of the absorption probabilities~$a_{i,\beta}(\lambda)$ in~\eqref{e:def_absorption_proba}, we deduce that, 
\begin{align}
  g_{b,\beta}(\lambda)
  =
  \begin{cases}
    0, & \text{if } \lambda \le \lambda_b^{(1)}(\beta),\\
    p_{b}^{(1)}, & \text{if } \lambda \in \left(\lambda_b^{(1)}(\beta), \lambda_b^{(2)}(\beta)\right),
  \end{cases}
\end{align}
and for any $\lambda > \lambda_b^{(1)}(\beta)$, $g_{b,\beta}(\lambda)$ is a convex combination of the probabilities $\left(p_i\right)_{i:\,\beta(i)=b}$, and thus~$p_b^{(1)}$ is its maximum value. Lemma~\ref{lem:g_b_decreasing} shows that $g_{b,\beta}$ is decreasing over~$\left(\lambda_b^{(1)}(\beta),+\infty\right)$. Consequently, $g_{\beta}(\lambda)=\sum_{b\in \cB} g_{b,\beta}(\lambda)$ can only increase when~$\lambda$ crosses one of the points~$\lambda_b^{(1)}(\beta)$, $b\in\cB$; between such points, $g$ is decreasing. With the restriction of~$\lambda$ being an integer, we deduce that the smallest value of $\lambda$ that maximizes $g$ lies in the set~$\Lambda(\beta)$. 

The near optimal eviction threshold~$\widehat{\lambda}^{*}$ is a maximizer of the function $\sum_{k=1}^{n_{samp}} \sum_{b\in \cB} g_{b,\beta_k}(\lambda)$ (see~\eqref{e:approx_expectation_g_beta} and~\eqref{e:lambdahatstar}). Using Lemma~\ref{lem:g_b_decreasing}, each function $g_{b,\beta_k}$ only increases at the discontinuity points $\lambda_{b}^{(1)}(\beta_k)$ and decreases afterwards. Thus, using the same arguments as in the proof of the first item of the theorem, we deduce the second item. This finishes the proof.


\end{proof}


Using Theorem~\ref{thm:lambda_star}, one can compute $\lambda^{*}(\beta)$ and $\widehat{\lambda}^{*}$ in $\cO(m_1 \, n_{\cI})$ and $\cO(m_1 \, n_{\cI} \, n_{\text{samp}})$ time, respectively. Indeed, Alg.~\ref{alg:gbeta_computation} evaluates $g_{\beta}$ for a fixed $\lambda$ in $\cO(n_{\cI})$ time, and $\lambda^{*}(\beta)$ is a maximizer of $g_{\beta}$ over the set $\Lambda(\beta)$ which has at most $m_1$ values. In practice, the cost can be substantially smaller because we only consider integer $\lambda$ and many candidates coincide across buckets and across samples. In particular, in the experiments of Section~\ref{sec:simulations}, we observe that $|\Lambda(\beta)|\leq 15$ even though the worst case bound is $m_1=200$. 







Next, we characterize the $\lambda^{*}(\beta)$ when $\bm{p}$ is the uniform distribution, denoted $\lambda^{*}_{\text{unif}}(\beta)$. This characterization provides insights on how the optimal eviction threshold depends on the problem parameters. It also enables to derive probabilistic upper bounds on $\widehat{\lambda}^{*}$. These bounds can be used to further support that the empirical observation $|\Lambda(\beta)|\ll m_1$ is not an artifact of the experiments in Section~\ref{sec:simulations}.

Lemma~\ref{lem:lambda_start_beta_unif} shows that $\lambda^{*}_{\text{unif}}(\beta)$ coincides with the maximum load in the classical \textit{balls and bins} process with~$n_{\cI}$ balls and~$m_1$ bins. It also shows that, among all distributions with support size equal to~$n_{\cI}$, $\lambda^{*}(\beta) \leq~ \lambda^{*}_{\text{unif}}(\beta)$.



\begin{lemmaSp}[Eviction threshold for uniform arrivals] 
\label{lem:lambda_start_beta_unif}
When $\bm{p}$ is the uniform distribution, the following holds, 
\begin{enumerate}
    \item The optimal eviction threshold for a fixed $\beta$ corresponds to the largest number of collisions in the heavy block, i.e., $\lambda^{*}_{\text{unif}}(\beta)= \max_{b\in \cB} n_{b}$. 

    \item $\lambda^{*}(\beta) \leq \lambda^{*}_{\text{unif}}(\beta)$.
 
\end{enumerate}
\end{lemmaSp}

\begin{proof}

When $\bm{p}$ is the uniform distribution, $w_{i,\beta}(\lambda)$ is constant across items from the same bucket, i.e., $w_{i,\beta}(\lambda)= w_{j,\beta}(\lambda)$ whenever $\beta(i)=\beta(j)$. It follows that, $a_{i,\beta}(\lambda)=\nicefrac{1}{n_{\beta(i)}}$, and thus,
\begin{align}
  g_{b,\beta}(\lambda)
  =
  \begin{cases}
    0, & \text{if } \lambda \le \lambda_b^{(1)}(\beta),\\
    \frac{1}{n_{\cI}}, & \text{if } \lambda \in \left(\lambda_b^{(1)}(\beta), +\infty\right).
  \end{cases}
\end{align}
Note that when $\bm{p}$ is uniform, $\lambda_{b}^{(1)}(\beta)=n_b -1$, and thus, $\lambda^{*}_{\text{unif}}(\beta)=\max_{b\in \cB} n_b$. Moreover, using Theorem~\ref{thm:lambda_star}, we can write, 
    \begin{align}\label{e:thm_unif_lambda_1}
        \lambda_b^{(1)}(\beta) =\frac{\sum_{j\in \cI_b} p_j}{p_b^{(1)}} - 1 \leq n_b -1 \implies  \lambda^{*}(\beta)  \leq \max_{b\in \cB } n_b=\lambda^{*}_{\text{unif}}(\beta).
    \end{align}
This finishes the proof. 

\end{proof}

Under Assumption~\ref{assum:IdealHash}, and in a regime where $n_{\cI}\gg m_1$ such that $m_{1}\to \infty$, combining Lemma~\ref{lem:lambda_start_beta_unif} with the sharp high-probability bounds on the maximum load in a balls and bins process~\cite[Thm. 1]{balls_bins_random_98}, justifies the following approximation,
\begin{align}
\lambda^{*}_{\text{unif}}(\beta) \approx 
  \frac{n_{\cI}}{m_1}
  + \Theta \left(\sqrt{\frac{n_{\cI} \ln(m_1)}{m_1}} \right).
  \label{eq:lambda_star_asymptotic_m1}
\end{align} 
Using Lemma~\ref{lem:lambda_start_beta_unif}, Theorem~\ref{thm:hp_lambda} derives then a probabilistic upper bounds on $\lambda^{*}(\beta)$ that hold for any $(m_1, n_{\cI})$. These bounds can then be used to derive probabilistic bounds, under Assumption~\ref{assum:IdealHash}, on $\widehat{\lambda}^{*}$, as follows,
\begin{align}
  \ProbaC{\{\beta_k\}_{k=1}^{n_{\text{samp}}}}{\widehat{\lambda}^{*} \leq x} 
  ~\geq 
  ~ \ProbaC{\{\beta_k\}_{k=1}^{n_{\text{samp}}}}{ \max_{k\in [n_{\text{samp}}]} \lambda^{*}(\beta_k) \leq x} ~=~ \left(\ProbaC{\beta}{\lambda^{*}(\beta) \leq x}\right)^{n_{\text{samp}}}. 
\end{align}



\begin{theoremSp}[High-probability upper bound on $\lambda^{*}(\beta)$]\label{thm:hp_lambda}
Under Assumptions~\ref{assum:IdealHash}, for any $m_1$, $n_{\cI}$, and $\delta\in (0,1)$, w.p. at least $1-\delta$, the following holds,
\begin{align} \label{eq:lambda_star_finite_m1}
    \lambda^{*}(\beta) \leq 
\frac{n_{\cI}}{m_1}
+ \sqrt{\frac{2 n_{\cI}}{m_1} \ln\left(\frac{m_1}{\delta}\right)}
+ \frac{1}{3}\ln\left(\frac{m_1}{\delta}\right).
\end{align} 
\end{theoremSp}
\begin{proof}
We show that the relation holds for $\lambda^{*}_{\text{unif}}(\beta)$.
 Under Assumption~\ref{assum:IdealHash}, $n_b$ is the sum of $n_{\cI}$ i.i.d. random variables that each indicates whether item $i$ hashes to bucket $b$ or not. Each of these random variables have a mean and variance equal to $\frac{1}{m_1}$ and $\frac{1}{m_1}(1-\frac{1}{m_1})\leq \frac{1}{m_1}$, respectively. Thus, applying \textit{Bernstein}'s inequality~\cite[Corollary 7.3]{Lecture_Bernstein_21} yields that, w.p. at least $1-\frac{\delta}{m_1}$, 
\begin{align}\label{e:thm_unif_lambda_2}
        n_b \leq \frac{n_{\cI}}{m_1} + \sqrt{\frac{2n_{\cI} \ln(m_1/\delta) }{m_1}} + \frac{1}{3}\ln(m_1/\delta). 
\end{align}
Moreover for any $x$, $\Proba{\max_{b\in \cB} n_b \leq x}  = 1 - \Proba{\exists b\in \cB:\; n_b \geq x } \geq  1 - m_1 \Proba{n_b\geq x}$.
From Lemma~\ref{lem:lambda_start_beta_unif}, we obtain the target result. This finishes the proof. 

 
\end{proof}

The high-probability bounds above on $\lambda^{*}(\beta)$ are distribution-agnostic and therefore do not capture how the skew of~$\bm p$ affects~$\lambda^{*}(\beta)$. A more $\bm p$-sensitive route is presented in Appendix~\ref{app:proba_bounds_lambda_star}.




\section{Simulations}
\label{sec:simulations}

\begin{figure}[t]
    \centering

    \begin{subfigure}[b]{0.32\textwidth}
        \centering
        \includegraphics[width=\linewidth]{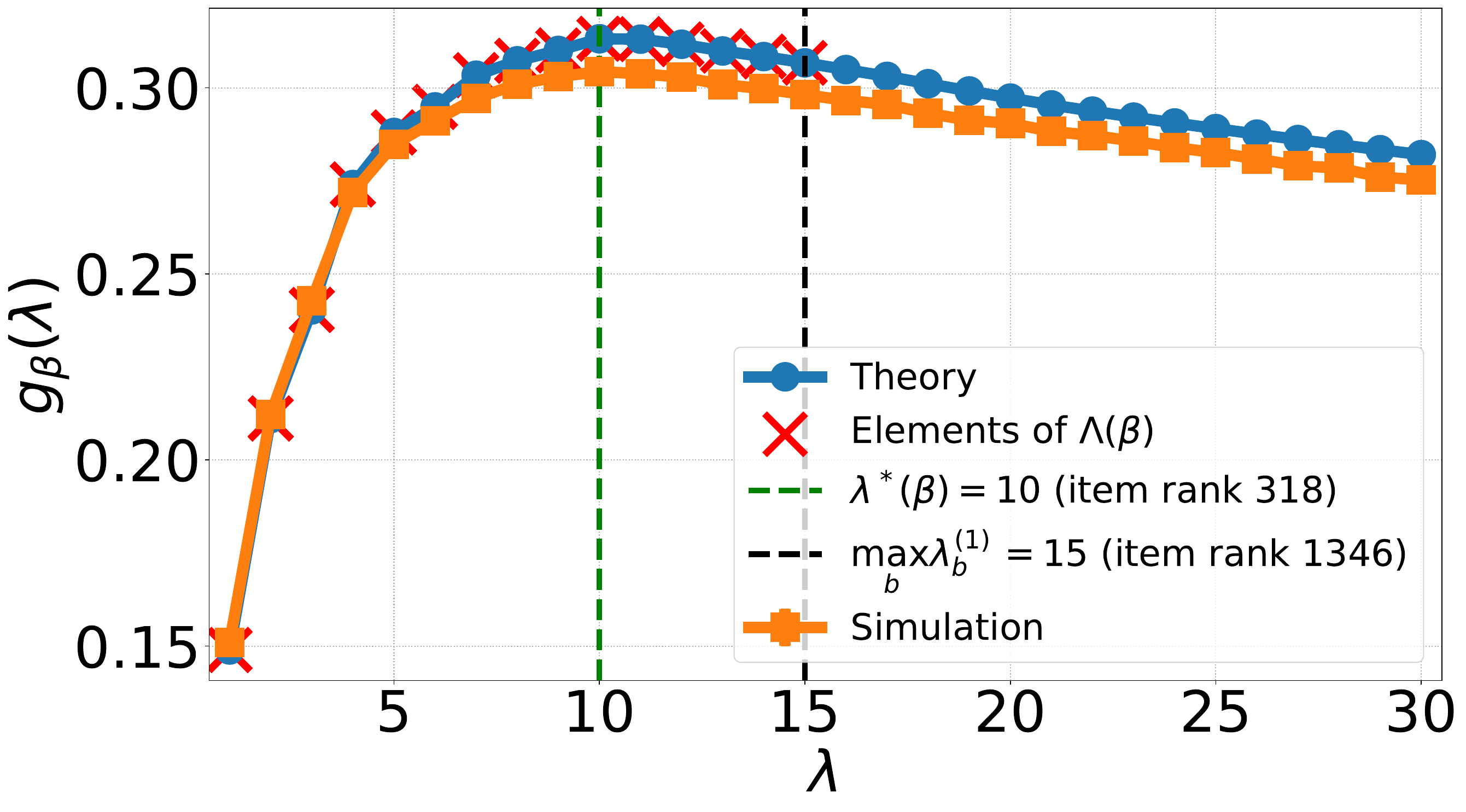}
        \caption{$\alpha = 0.8$}
        \label{fig:g_lambda_08}
    \end{subfigure}
    \hfill
    \begin{subfigure}[b]{0.32\textwidth}
        \centering
        \includegraphics[width=\linewidth]{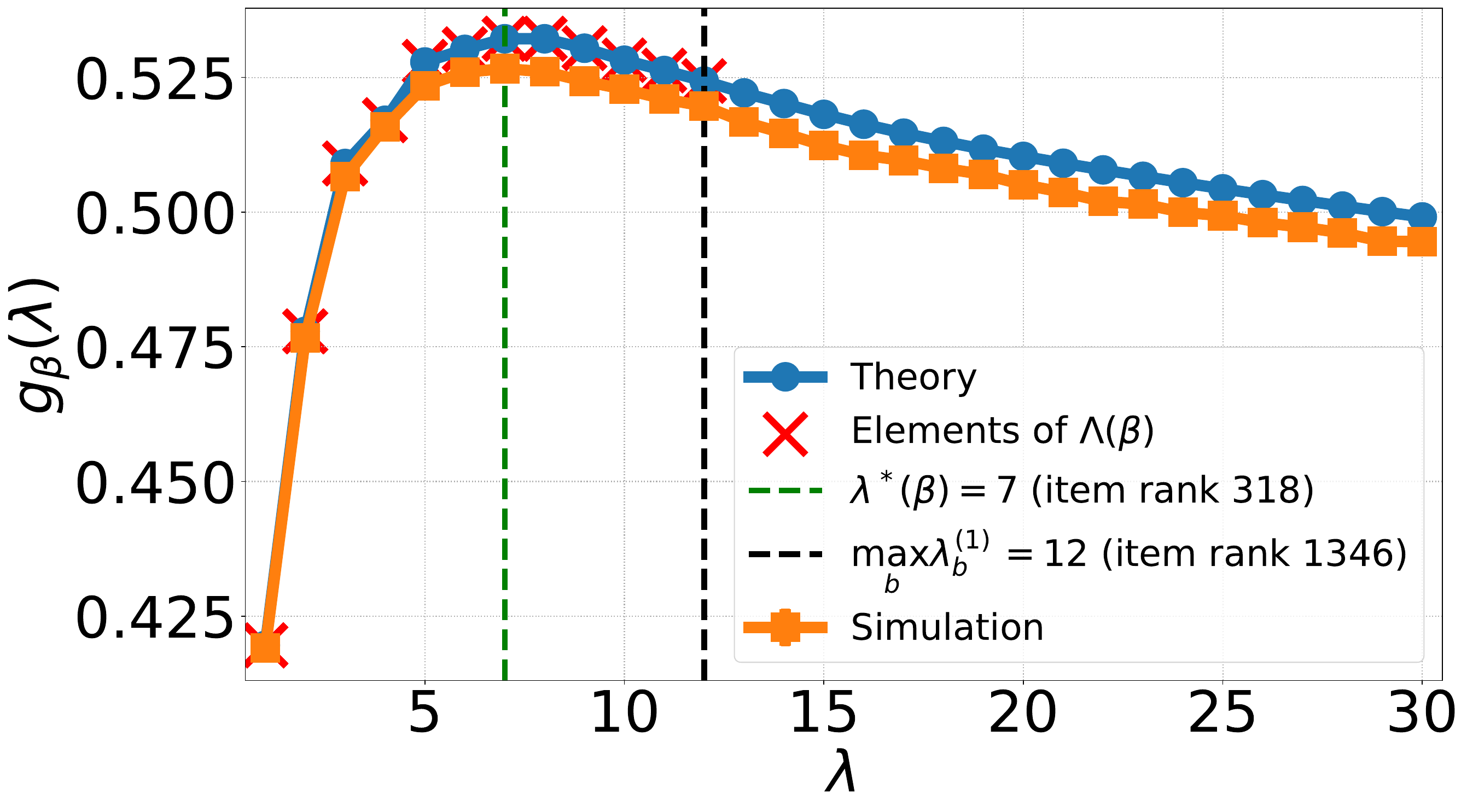}
        \caption{$\alpha = 1.0$}
        \label{fig:g_lambda_10}
    \end{subfigure}
    \hfill
    \begin{subfigure}[b]{0.32\textwidth}
        \centering
        \includegraphics[width=\linewidth]{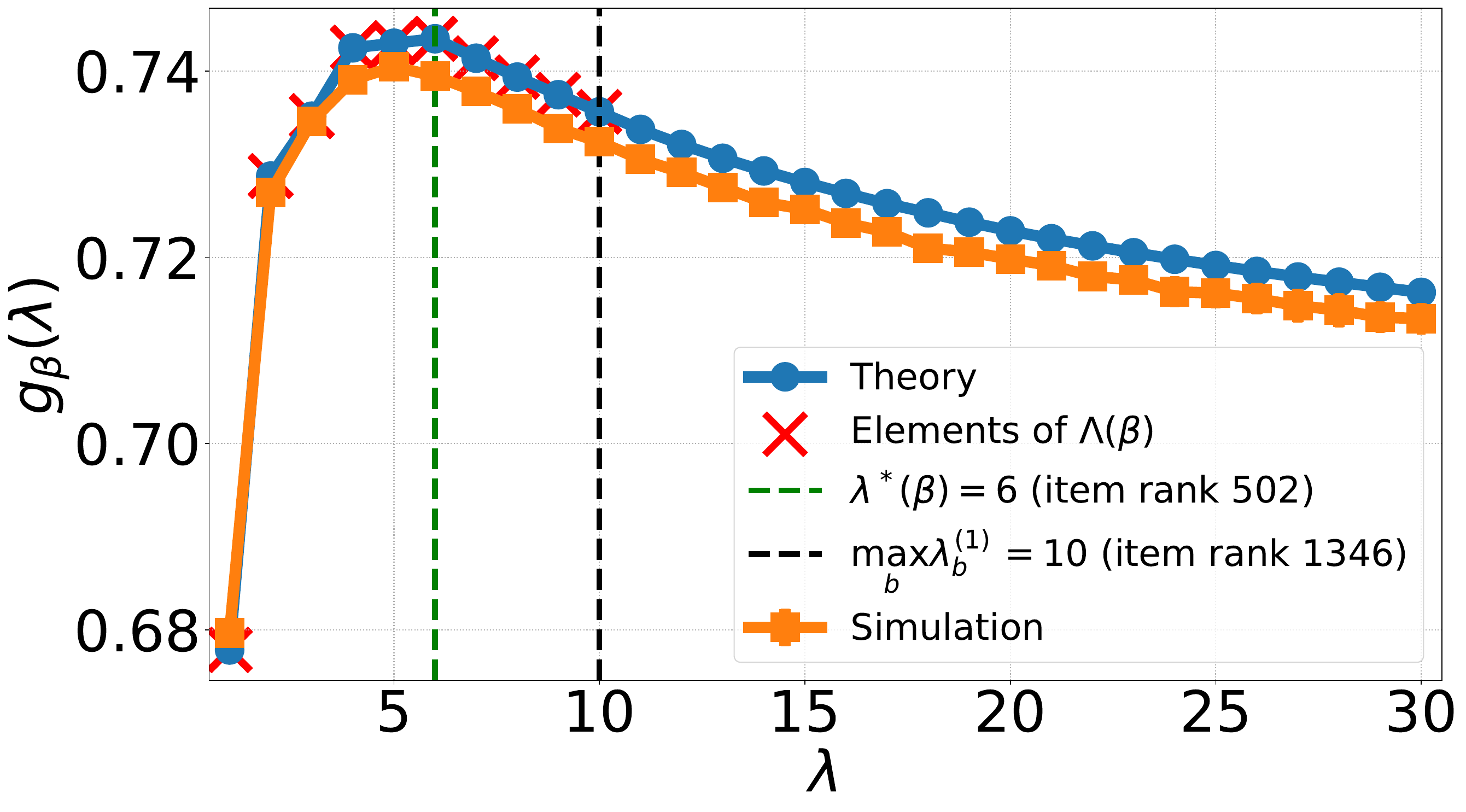}
        \caption{$\alpha = 1.2$}
        \label{fig:g_lambda_12}
    \end{subfigure}

    \caption{Estimation of $\E{\overline{V_{\cB}}(\tau)}$ via $g_{\beta}(\lambda)$ for Zipf request distributions with different skew parameters $\alpha$, $n_{\cI}=10^{4}$, $\tau=5\times 10^{5}$, $n_{\text{runs}}=100$, $m_1=200$.}
    \label{fig:g_lambda_zipf_alpha}
\end{figure}


In this section, we validate our results on arrivals from the Zipf distribution with skew parameter~$\alpha$, i.e., $\cI =[n_{\cI}]$ and $p_{i}\propto\nicefrac{1}{i^{\alpha}}$. In practice, $\ElaSk$ is run on finite streams, or over sliding windows. We therefore evaluate how accurately the asymptotic characterizations in Theorem~\ref{thm:counters_limit} and Corollary~\ref{cor:limting_error_expectation} predict the finite-$\tau$ expected error. We also assess how well the eviction threshold tuned to minimize the asymptotic expected error performs when the objective is the finite time expected error. 



Similarly to Corollary~\ref{cor:limting_error_expectation}, one can use Lemma~\ref{lem:Count_ElasticSketch_Finite} (Section~\ref{ss:proof_limit_V_and_Y}) to obtain a finite time expression for the expected error,
    \begin{align}
         \frac{1}{\tau}   \Esub{\bm{h}}{\text{Err}_{(0)}(\tau)} = \frac{1}{m_2} \left( 1- \E{\overline{V_{\cB}}(\tau)} \right): \; \; \overline{V_{\cB}}(\tau) \triangleq \frac{1}{\tau}\sum_{b\in\cB} V_b^{+}(\tau). 
    \end{align}
Hence, minimizing the finite time expected error is equivalent to maximizing $\overline{V_{\cB}}(\tau)$. We therefore validate our approach through the metric~$\overline{V_{\cB}}(\tau)$.


Figure~\ref{fig:g_lambda_zipf_alpha} reports estimates of $\E{\overline{V_{\cB}}(\tau)}$ as a function of~$\lambda$ for the same hash function $\beta$. The curve labeled "\textbf{Simulations}" is obtained by averaging $\overline{V_{\cB}}(\tau)$ over $n_{\text{runs}}=100$ independent streams drawn form a Zipf distribution with skew parameter $\alpha\in \{0.8, 1.0, 1.2\}$. The curve labeled "\textbf{Theory}" corresponds to the asymptotic prediction $\E{\overline{V_{\cB}}(\tau)}\approx g_{\beta}(\lambda)$ provided by Theorem~\ref{thm:counters_limit} (for large $\tau$). The figure also shows the candidates set $\Lambda(\beta)$ from Theorem~\ref{thm:lambda_star}. In particular, it highlights as vertical lines (i) the maximizer of $g_{\beta}(\lambda)$ over $\Lambda(\beta)$, namely $\lambda^{*}(\beta)$, and (ii) the largest element of $\Lambda(\beta)$, i.e., $\max_{b\in \cB} \lambda_{b}^{(1)}(\beta)$.
Both quantities are induced by a specific bucket~$b$; we additionally report the rank of the highest-probability item among those hashing to~$b$.

Even for a finite time stream length $\tau=5\times 10^{5}$ with $n_{\cI}=10^{4}$, the asymptotic prediction $g_{\beta}(\lambda)$ closely tracks the Monte Carlo estimate of $\E{\overline{V_{\cB}}(\tau)}$. Moreover, evaluating $g_{\beta}(\lambda)$ is substantially cheaper. It can be computed in $\cO(n_{\cI})$ for a fixed $(\lambda,\beta)$, via Alg.~\ref{alg:gbeta_computation}, whereas simulations require $\cO(n_{\text{runs}}\, \tau)$. We also observe that in general the theoretical curve is slightly above the empirical estimates.

Although the worst-case size of~$\Lambda(\beta)$ is at most~$m_1=200$, in this experiment it is much smaller: $|\Lambda(\beta)|=15$, $12$, and $10$ for~$\alpha\in \{ 0.8, 1.0, 1.2 \}$, respectively. This reduction is mainly due to the restriction of~$\lambda$ being an integer, which makes the maximization over~$\Lambda(\beta)$ extremely fast. Moreover, for the three values of~$\alpha$, the value of~$\lambda$ that maximizes the empirical estimate coincides with the asymptotically optimal threshold~$\lambda^{*}(\beta)$.

Finally, the function $g_{\beta}(\lambda)$ varies sharply for small $\lambda$, but becomes comparatively flat after the optimum and then decreases gradually as $\lambda$ grows. This suggests that, in practice, one may replace the exact optimizer by a simpler, tractable surrogate such as $\lambda_{\text{ub}}(\beta)\triangleq \max_{b\in \cB} \lambda_{b}^{(1)}(\beta)$. Indeed, for the considered experiments, $g_{\beta}(\lambda^{*}(\beta))\approx g_{\beta}(\lambda_{\text{ub}}(\beta))$. Moreover, as shown in \S~\ref{ss:optimal_eviction_threshold} and Appendix~\ref{app:proba_bounds_lambda_star}, $\lambda_{\text{ub}}$ is amenable to probabilistic analysis accounting for the skewness of $\bm{p}$. Nonetheless, the apparent flatness of $g_{\beta}(\lambda)$ beyond the optimum can be misleading: while the average changes little, large values of $\lambda$ can induce substantial variability in performance, as observed in Fig.~\ref{fig:are_n_runs100_m150_nItems20000_T500000_widthCMS_200_a12} of the introduction.

Using "xxHash" hash functions, and generating $\cup_{k=1}^{n_{\text{samp}}} \Lambda(\beta_k)$, with $n_{\text{samp}}=1000$, we observe only $27$, $25$, and $22$ distinct elements for $\alpha\in \{0.8, 1.0, 1.2\}$, respectively. This confirms that, in practice, the near-optimal eviction threshold $\widehat{\lambda}^{*}$ can be computed efficiently.


\section{Proof of Theorem~\ref{thm:counters_limit}} 
\label{ss:Proof_counters_limit}


  
Section ~\ref{ss:Def_not_thm_counters_limit} introduces additional definitions and notation used in the proof. Section~\ref{ss:proof_s_limit} proves item~\ref{item:thm_limit_S} of Theorem~\ref{thm:counters_limit}, and Section~\ref{ss:proof_limit_V_and_Y} proves items~\ref{item:thm_limit_V} and~\ref{item:thm_limit_Y} of the theorem. 

\subsection{Definitions and Notation}
\label{ss:Def_not_thm_counters_limit}

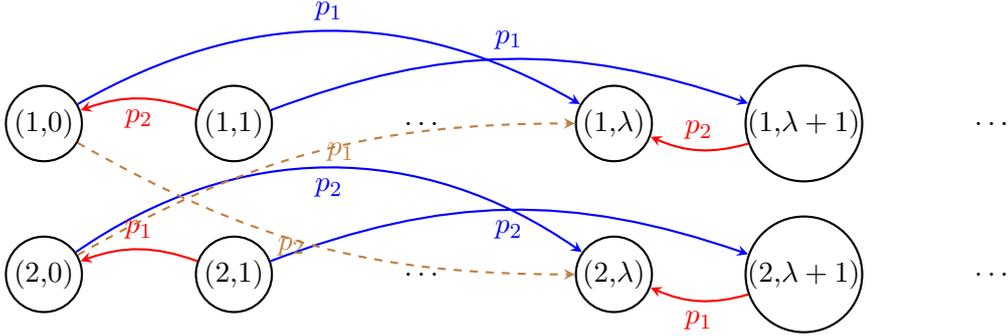
\begin{figure}[h]
\centering 
\begin{tikzpicture}[->, >=stealth, auto, node distance=2.5cm, on grid]
    \tikzstyle{state} = [circle, draw, thick, minimum size=10mm, inner sep=0pt]

    \tikzstyle{dec}      = [red,   thick]                 
    \tikzstyle{incsame}  = [blue,  thick]                 
    \tikzstyle{incother} = [brown, thick, dashed]         

    \node[state] (N1_0)                    {(1,0)};
    \node[state] (N1_1)  [right=of N1_0]   {(1,1)};
    \node         (N1dots) [right=of N1_1] {$\dots$};
    \node[state] (N1_lam)  [right=of N1dots]  {(1,$\lambda$)};
    \node[state] (N1_lam1) [right=of N1_lam]  {(1,$\lambda+1$)};
    \node         (N1dots2)[right=of N1_lam1] {$\dots$};

    \node[state] (N2_0)  [below=2cm of N1_0] {(2,0)};
    \node[state] (N2_1)  [below=2cm of N1_1] {(2,1)};
    \node         (N2dots)  [below=2cm of N1dots]  {$\dots$};
    \node[state] (N2_lam)  [below=2cm of N1_lam]  {(2,$\lambda$)};
    \node[state] (N2_lam1) [below=2cm of N1_lam1] {(2,$\lambda+1$)};
    \node         (N2dots2)[below=2cm of N1dots2] {$\dots$};

    \draw[incsame] (N1_1)    edge[bend left=20]  node[above] {$p_1$} (N1_lam1);
    \draw[incsame] (N1_0)    edge[bend left=30]  node[above] {$p_1$} (N1_lam);

    \draw[dec]     (N1_lam1) edge[bend left=20]  node[above] {$p_2$} (N1_lam);
    \draw[dec]     (N1_1)    edge[bend right=20] node[below] {$p_2$} (N1_0);

    \draw[incsame] (N2_1)    edge[bend left=20]  node[below] {$p_2$} (N2_lam1);
    \draw[incsame] (N2_0)    edge[bend left=35]  node[below] {$p_2$} (N2_lam);

    \draw[dec]     (N2_lam1) edge[bend left=20]  node[below] {$p_1$} (N2_lam);
    \draw[dec]     (N2_1)    edge[bend right=20] node[above] {$p_1$} (N2_0);

    \draw[incother] (N2_0)   edge[bend left=15]  node[right] {$p_1$} (N1_lam);
    \draw[incother] (N1_0)   edge[bend right=15] node[left]  {$p_2$} (N2_lam);
\end{tikzpicture}
    \caption{Illustration of the Markov chain $M_b$ when $n_b=2$.}
    \label{fig:markov_chain} 
\end{figure}

Define $\mathcal{T}_b$ be the set of eviction times in bucket $b$, i.e.,
\begin{align}
    \mathcal{T}_b \triangleq \{ t\in \mathbb{N}: \;  S_b(t) \neq S_b(t+1)  \},
\end{align}
and $T_b^{\text{evict}}(t)$ is the last eviction time at step $t$, i.e., 
\begin{align}
       T_b^{\text{evict}}(0) = 0 , \; T_b^{\text{evict}}(t) \triangleq \max \left ( \mathcal{T}_b \cap \{ 0, 1, \dots , t-1 \} \right), \; \forall t \geq 1.  
\end{align}

\noindent To study the limiting behavior of $\bm{S}$, we define the stochastic process $M_b$ for each bucket $b$ as,
\begin{align}
M_b(t)\triangleq (S_b(t),U_b(t)), \; 
U_b(t)\triangleq \lambda V^{+}_b(t)-V^{-}_b(t).
\end{align}
The score $U_b(t)$ governs updates of $S_b(t)$ (see line~\ref{line:updates_S} in Alg.~\ref{Alg:Elastic_Sketch}). Conditioned on the hash functions $\{h_{\ell}\}_{\ell=1}^{d}$ and $\beta$, the evolution of $M_b$ is determined solely by the stream $\bm{R}$. Hence, under Assumption~\ref{assum:stationary_stream}, $M_b$ is a time-homogeneous discrete time Markov chain on $\cI_b \cup \{-1\}\times\mathbb{N}$. Initially, $M_b(0)=(-1,0)$, and the one step transitions are given by, 
\begin{align}\label{e:Markov_Block1}
(S_b(t+1), U_b(t+1)) =
\begin{cases}
\left(S_b(t), U_b(t) \right), &\text{w.p. } 1-\mu_{b},  \\
\left(S_b(t), U_b(t) + \lambda\right), &\text{w.p. } p_{S_b(t)}, \text{ if } U_b(t) >0,  \\
\left(S_b(t), U_b(t) - 1\right), &\text{w.p. }  \mu_{b} - p_{S_b(t)},  \text{ if } U_b(t) >0,\\
\left(i, \lambda\right), &\text{w.p. }  p_i, \text{ if } U_b(t) = 0, \; \forall i \in \cI_b,
\end{cases}
\end{align}
In words, upon a request for an item not in the bucket in question, $M_b(t)$ remains unchanged (w.p. $1-\mu_{b}$). Moreover, when $U_b(t)$ is positive, $S_b(t)$ remains unchanged, and either $U_b(t)$ increases by $\lambda$ (with probability $p_{S_b(t)}$) or decreases by $1$ (w.p. $\mu_{b} - p_{S_b(t)}$). Once $U_b(t)$ hits $0$, $S_b(t)$ may switch to a new item $i$ w.p. $p_i$ and $U_b$ increases by $\lambda$ anyway. Figure~\ref{fig:markov_chain} illustrates the one step transitions of this Markov process when the total number of items is $2$ and the number of buckets is $1$. We denote the set of states with $U_b=0$ as $\overline{\mathbf 0_b}$, i.e., $\overline{\mathbf{0}_b} \triangleq\{ (i,0): \; i\in \cI_b\cup\{-1\} \}$.

\subsection{Proof of item~\ref{item:thm_limit_S} in Theorem.~\ref{thm:counters_limit}}
\label{ss:proof_s_limit}

We prove that, for any bucket $b\in \cB^{-}_{\beta}(\lambda)$, $(S_b(t))_t$ switches state infinitely often, and for any bucket $b\in \cB^{+}_{\beta}(\lambda)$, $S_b(t)$ converges a.s. in $t$, such that for any item $i\in \cI_b$, $\Proba{S_{b}^{\infty}=i}=  a_{i,\beta}(\lambda)$. 
For any bucket $b\in \cB_{\beta}^{0}(\lambda)$, $S_{b}(t)=-1$ at any step $t$. Lemma~\ref{lem:Relation_S_M} shows that the transience/recurrence of $M_b$ determines the limiting behavior of $S_b$; $S_b$ converges and $\lim_{t\to\infty} T_b^{\text{evict}}(t)/t=0$ a.s. when $M_b$ is transient, otherwise, $S_b$ switches state infinitely often and $\lim_{t\to \infty} (t-T_b^{\text{evict}}(t))/t=0$ a.s.. Furthermore, Claim~\ref{lem:Transience_M} shows that for any bucket $b\in \cB_{\beta}^{+}(\lambda)$ ($\lambda > \lambda_{b}^{(1)}(\beta)$), the Markov chain $M_{b}$ is transient, and it is recurrent for any bucket $b^{-}\in \cB_{\beta}^{-}(\lambda)$ ($\lambda \leq \lambda_{b}^{(1)}(\beta)$). Combining these two results, we deduce that, for any bucket~$b\in \cB^{-}_{\beta}(\lambda)$, $(S_b(t))_t$ switches state infinitely often, and for any bucket~$b\in \cB^{+}_{\beta}(\lambda)$, $S_b(t)$ converges a.s. in $t$. The distribution of $S_b^{\infty}$ follows directly from Lemma~\ref{lem:Proba_S_infinity}. This proves item~\ref{item:thm_limit_S} of Thm.~\ref{thm:counters_limit}. Below, we present Lemmas~\ref{lem:Relation_S_M} and~\ref{lem:Proba_S_infinity}, and Claim~\ref{lem:Transience_M}, along with their proofs.

\begin{lemmaSp}[Relation between $M_b$ and $S_b$]\label{lem:Relation_S_M}
Under Assumption~\ref{assum:stationary_stream}, the transience/positive recurrence of $M_b$, for any $b\in \cB_{\beta}^{+}(\lambda)\cup \cB_{\beta}^{-}(\lambda)$, determines the limiting behavior of $S_b$: 
\begin{itemize}
    \item If $M_b$ is recurrent, then $(S_b(t))_t$ switches state infinitely often ($|\mathcal{T}_b|=\infty$), and $\\$ 
    $\lim_{t\to \infty} \frac{T_b^{\text{evict}}(t)}{t}=~1$, a.s.. 

    \item If $M_b$ is transient, then $(S_b(t))_t$ converges ($|\mathcal{T}_b|$ is finite), and $\lim_{t\to\infty} \frac{T_b^{\text{evict}}(t)}{t}=0$, a.s.. 
\end{itemize}

\end{lemmaSp}

\begin{proof}[Proof of Lemma~\ref{lem:Relation_S_M}]




Define~$\mathcal{Z}_b$ as the set of time instants where $M_b$ visits $\overline{\bm{0}}_b$, i.e., 

    \begin{align}
            \mathcal{Z}_b \triangleq \left\{ t\in \mathbb{N}: \; U_b(t)=0 \right\}.
    \end{align}

\noindent \textbf{$M_b$ is transient.} It follows that $ \mathcal{Z}_b$ is finite almost surely (a.s.). Moreover, $ \mathcal{T}_b \subset  \mathcal{Z}_b$, because $S_b$ can only change value when $M_b$ is in $\overline{\bm{0}}_b$ (see~\eqref{e:Markov_Block1}). It follows then that $ \mathcal{T}_b$ is finite a.s.. Writing $T^{\text{evict}}_b(t) = \max  \mathcal{T}_b \cap \{0, \ldots, t-1 \}$, we obtain the almost sure limit $\lim_{t\to\infty} T^{\text{evict}}_b(t)=t^{*}$ with $t^{*}=\max  \mathcal{T}_b$ if $ \mathcal{T}_b \neq \emptyset$ and it is equal to $0$ otherwise. It follows that $S_b$ stabilizes a.s. after $t^{*}+1$, i.e., $S_b(t)= S_b(t^{*}+1)$, for all $t\geq t^{*}+1$, and $\lim_{t\to \infty} T^{\text{evict}}_b(t)/t=0$, a.s..

\paragraph{$M_b$ is positive recurrent.}
Since $M_b$ is positive recurrent, the set $ \mathcal{Z}_b$ of return times is infinite a.s.; write $ \mathcal{Z}_b=(z_i)_{i\in\mathbb{N}}$ with $z_i<z_{i+1}$. By the strong Markov property, the increments $(z_{n+1}-z_n)_{n\in\mathbb{N}}$ are i.i.d. with finite mean. 
Next, we show that $ \mathcal{T}_b$ is also infinite and write $ \mathcal{T}_b=(t_i)_{i\in\mathbb{N}}$. We also show that there exist identically distributed random variables $(X_n)_{n\in\mathbb{N}}$ with $\E{X_0}<\infty$ such that
\begin{align}\label{e:finite_difference_T}
  t_{n+1}-t_n \le X_n \quad \text{a.s.\ for all } n.
\end{align}
For $t\ge 0$, let $l(t)$ be the unique index such that $t_{l(t)}\le t< t_{l(t)+1}$. Then $T^{\text{evict}}_b(t)=t_{l(t)}$ and
\begin{align}
  0 \le t-T^{\text{evict}}_b(t) < t_{l(t)+1}-t_{l(t)} \le X_{l(t)} .
\end{align}
Since $X_{l(t)} \stackrel{d}{=} X_0$ with $\E{X_0}<\infty$ and $t\to\infty$, we conclude that $\lim_{t\to\infty}(t-T^{\text{evict}}_b(t))/t=0$.

\noindent\textbf{Proof that $ \mathcal{T}_b$ is infinite and \eqref{e:finite_difference_T} holds.}
We represent $ \mathcal{T}_b$ by thinning the renewal process $ \mathcal{Z}_b$ with i.i.d. uniform random variables. 
Let $\{u_{z}\}_{z\in \mathcal{Z}_b}$ be i.i.d. $\mathrm{Unif}[0,1]$, independent of $ \mathcal{Z}_b$, and define
\begin{align}
   \mathcal{T}_b = \left\{ z\in \mathcal{Z}_b : u_{z} \le 1 - p_{S(z)} \right\}.
\end{align}
This is valid because, whenever $U_b=0$, the probability that $S_b$ changes state at the next step is $1-p_{S_b}$ (see~\eqref{e:Markov_Block1}). 
Recall that $p^{(1)}_b\triangleq \max_{i\in \cI_{b}} p_i$  and define the dominated thinning
\begin{align}
  \mathcal{T}' \triangleq \bigl\{\, z\in \mathcal{Z}_b : u_{z} \le 1 - p^{(1)}_b \,\bigr\}.
\end{align}
Clearly $\mathcal{T}' \subseteq  \mathcal{T}_b$. Writing $\mathcal{T}'=(t'_n)_{n\in\mathbb{N}}$, the inter-arrivals $(t'_{n+1}-t'_n)$ are i.i.d., and $t^{'}_1-t^{'}_0\stackrel{d}{=} \sum_{k=1}^{G_n} (z_{k}-z_{k-1})$, where $G_n\sim \mathrm{Geom}(1-p^{(1)}_b)$ on $\{1,2,\dots\}$, and $(z_k-z_{k-1})$ are i.i.d., independent of $G_n$. Hence

\begin{align}
  \E{t'_{n+1}-t'_n} = \E{G_0} \E{z_{1}-z_0} = \frac{1}{1-p^{(1)}_b}\E{z_{1}-z_0} < \infty,
\end{align}
so $\mathcal{T}'$ is infinite a.s. and has i.i.d. inter-arrivals with finite mean.

Since $\mathcal{T}'\subseteq \mathcal{T}_b$, every inter-arrival in $ \mathcal{T}_b$ is bounded by a (possibly larger) enclosing inter-arrival of $\mathcal{T}'$: for each $n$, there exists an index $j(n)$ with $t_{n+1}-t_n \le t'_{j(n)+1}-t'_{j(n)}$. Set $X_n \triangleq t'_{j(n)+1}-t'_{j(n)}$. Then $(X_n)$ are identically distributed as $t'_{1}-t'_{0}$ and satisfy \eqref{e:finite_difference_T} with $\E{X_0}<\infty$.

\noindent \textbf{$M_b$ is null recurrent.} The reasoning used to prove that~$ \mathcal{T}_b$ is infinite when~$M_b$ is positive recurrent applies here as well, but the inter-arrivals have infinite mean. Nonetheless, they are finite a.s., and thus, $X_0/t$ is $0$, allowing to deduce that $(t-T^{\text{evict}}_b(t))/t$ is $0$ as well. This finishes the proof.


\end{proof}

\begin{claimSp}[Transience/recurrence of $M_b$]\label{lem:Transience_M}
    The comparison between $\lambda$ and $\lambda_b^{(1)}(\beta)$ determines the transience/positive recurrence of $M_b$: 

\begin{itemize}

       \item If $b\in \cB_{\beta}^{-}(\lambda)$, i.e., $\lambda \leq \lambda_{b}^{(1)}(\beta)$, then $(M_b(t))_t$ is recurrent.
    \item If $b\in \cB_{\beta}^{+}(\lambda)$, i.e., $\lambda > \lambda_{b}^{(1)}(\beta)$, then $(M_b(t))_t$ is transient.

\end{itemize}

\end{claimSp}

\begin{lemmaSp}[Probability distribution of $S_b^{\infty}$] 
\label{lem:Proba_S_infinity}
For any bucket $b\in \cB^{+}_{\beta}(\lambda)$, and for any item $i\in \cI_b$, $\Proba{S_{b}^{\infty}=i}=  a_{i,\beta}(\lambda)$, such that the function $a_{i,\beta}$ is defined in~\eqref{e:def_absorption_proba}. 
\end{lemmaSp}

To prove Claim~\ref{lem:Transience_M} and Lemma~\ref{lem:Proba_S_infinity}, we define the stochastic process~$G$ on~$\mathbb{N}$ as follows,
\begin{align}\label{e:law_Markov_G}
    G(t+1) = \begin{cases}
    G(t), &\text{ w.p. } 1-\mu, \\ 
    \max \left(G(t)-1 , 0\right), &\text{ w.p. } \mu - p, \\ 
    G(t) + \lambda,  &\text{ w.p. } p. 
    \end{cases}
\end{align}
and $0<p \leq \mu \leq 1$. The process~$G$ captures the evolution of~$M_b$ for a fixed value of~$U_b$. Define $r_n$ as the return probability of $G$ to $0$ when starting at $n>0$, and $T_a$ as the hitting time of $G$ on $a\geq 0$, 
\begin{align}\label{e:def_rn_G}
&r_n\triangleq \Proba{T_0 < \infty \mid G(0) =n},  \\ 
&T_a \triangleq \inf\{t\geq 0:\; G(t)=a\},\qquad a\in\mathbb{N}. \label{e:Hitting_time_G_def}
\end{align}

\begin{claimSp}[Hitting probabilities of $G$: Exact expression]
\label{cor:Return_to_0_Proba_G}
For all $n\in\mathbb{N}\setminus \{0\}$, $r_n=(r_1)^n$. Moreover, 
\begin{align}
        r_1 = \begin{cases}
                    1, &\text{ if } \lambda\geq \nicefrac{\mu}{p} -1 \\
                    \text{Unique root in } (0,1) \text{ of } \phi(\cdot, \lambda)- \mu/p, & \text{otherwise.}
        \end{cases}
\end{align}
\end{claimSp}

\begin{proof}[Proof of Claim~\ref{cor:Return_to_0_Proba_G}]

Note that for any $n\geq 1$, $\E{G(1)-G(0)\mid G(0)=n}= p\, (\lambda +1) - \mu$, which is strictly negative whenever $\frac{\mu}{p} > \lambda + 1$. Using a \textit{Foster's criterion}~\cite[Thm. 114]{serfozo2009basics}, we deduce that $G$ is positive recurrent, and thus $r_n=1$ for any $n$ in this case.

To derive $r_n$ for other values of $\lambda$, we prove that, 

\begin{align} \label{e:multiplicative_rn}
        \forall n,m\ge 1, \;  r_{n+m}=r_n  r_m.
\end{align}

\noindent This implies that $r_n=r_1^n$. Moreover, from the Markov property, we obtain, 
\begin{align}\label{e:R_rec_raw}
r_n =
\begin{cases}
(1-\mu) r_1 + p r_{1+\lambda} + (\mu - p), & \text{ if } n=1, \\
(1-\mu) r_n + p r_{n+\lambda} + (\mu-p) r_{n-1}, &\text{ if } n\geq 2.
\end{cases}
\end{align}
Developing~\eqref{e:R_rec_raw}, we obtain that $r_1$ must be a root in $[0,1]$ of the function $f$ given by,
\begin{align}
f(x) \triangleq px^{\lambda+1} -\mu x + \mu-p= (x-1) \left(p\phi(x,\lambda)- \mu)\right),
\end{align}

\noindent where $\phi(x,\lambda) = \sum_{k=0}^{\lambda} x^{k}$. For $x\in [0,1]$, $\phi(x,\lambda)$ takes values in $[0,\lambda+1]$, and it is strictly increasing in $x$.
 Thus, (i) when $\lambda\leq \frac{\mu}{p} -1$, $1$ is the unique root of $f$, and therefore $r_n=1$, $\forall n$, i.e., $G$ is recurrent, and (ii), when $\lambda>\frac{\mu}{p} -1$, $f$ has two distinct roots,~$1$, and~$x_0\in (0,1)$, such that $\phi(x_0,\lambda)=\mu/p$. Next, we prove that $r_1<1$ when $\frac{\mu}{p} < \lambda + 1$, to deduce that $r_1$ is the solution to the equation $\phi(\cdot,\lambda)=\mu/p$.  

 Below, we prove \eqref{e:multiplicative_rn} and then that $r_1<1$ when $\lambda>\frac{\mu}{p} -1$.

\noindent \textbf{Proof of \eqref{e:multiplicative_rn}.} Because downward moves in $G$ are \emph{skip-free} (only $-1$), any path that reaches $0$ from $n+m$ must visit $m$ first; hence
$
\{T_0<\infty\}\subseteq\{T_m<\infty\},
$
 (see~\eqref{e:Hitting_time_G_def} for the definition of $T_a$ for $a\in \mathbb{N}$) and therefore
\begin{align}
r_{n+m}
&= \Proba{T_0<\infty\mid G(0)=n+m} \\ \label{e:lem_R_multi_G_1}
&= \Proba{T_0<\infty \mid T_m<\infty,\; G(0)=n+m }\Proba{T_m<\infty \mid G(0)=n+m},
\end{align}
since $\Proba{T_0<\infty \mid T_m=\infty, \; G(0)=n+m}=0$.

On $\{T_m<\infty\}$ and $G(0)=n+m$, we have $T_0>T_m$ and thus
$
T_0=\inf\{t\ge 0:\, G(t+T_m)=0\}.
$
Because $T_m$ is a stopping time, the strong Markov property gives that
$
(G(t+T_m))_{t\ge 0}|\{G(0)=n+m, T_m<\infty\})
$
has the same law as
$
(G(t))_{t\ge 0}|\{G(0)=m\}).
$
Hence
\begin{align}\label{e:lem_R_multi_G_2}
\Proba{T_0<\infty \mid T_m<\infty, \; G(0)=n+m}
= \Proba{T_0<\infty \mid G(0)=m}
= r_m.
\end{align}
By a space-homogeneity argument, we deduce that
$
\Proba{T_m<\infty \mid G(0)=n+m}=r_n.
$
Combining this with~\eqref{e:lem_R_multi_G_1} and~\eqref{e:lem_R_multi_G_2} finishes the proof.

\noindent \textbf{Proof that $ r_1 < 1$ when $ \frac{\mu}{p} < \lambda + 1$.} Let $\{J(t)\}_{t\ge 0}$ be the $\mathbb{Z}$-valued random walk with i.i.d. increments $(\zeta_j)_{j\geq 1}$ such that, 
\begin{align}
\Proba{\zeta_1=\lambda}=p,\qquad
\Proba{\zeta_1=-1}=\mu-p,\qquad
\Proba{\zeta_1=0}=1-\mu,
\end{align}
and $J(t)=n+\sum_{s=1}^t \zeta_s$ for some $n>0$. Let $\mu$ be the expectation of $\zeta_1$, i.e., $\gamma= p(\lambda+1)-\mu$. We also denote the moment generating function of $\zeta_1$ as $\psi(\theta)$, i.e.,  
\begin{align}
\psi(\theta)\triangleq \E{e^{\theta \zeta_1}}=p e^{\theta \lambda}+(\mu-p) e^{-\theta}+(1-\mu).
\end{align}
When $\frac{\mu}{p} < \lambda + 1$, $\psi^{'}(0)=\mu>0$. And since $\psi(0)=1$, then there exists $\theta_0>0$ such that $\psi(-\theta_0)\in (0,1)$. 

When $G(0)=n$, $G$ and $J$ visit $0$ for the first time at the same time, because $G(t)= \max\left( J(t), 0 \right)$ (see \eqref{e:law_Markov_G}). Thus, we can bound $r_n$ as follows,  
\begin{align}
r_n
=\Proba{\exists t\ge 0:\, J(t) =0}
\le \sum_{t=0}^\infty \Proba{J(t)=0}
\le \sum_{t=0}^\infty \E{e^{-\theta_0 J(t)}} 
= \sum_{t=0}^\infty e^{-\theta_0 n}\,\left(\psi(-\theta_0)\right)^t
= \frac{e^{-\theta_0 n}}{1-\psi(-\theta_0)}.
\end{align}
It follows that there exists $n_0\geq 1$ such that $r_{n_0}<1$, which means that~$G$ is transient, and thus~$r_n<1$ for any $n\geq1$ when~$\frac{P}{p}<\lambda +1$. We deduce then that~$r_1$ is the unique solution to the equation~$\phi(\cdot ,\lambda)$ in the interval $(0,1)$, which finishes the proof.

\end{proof}

Now, we are ready to leverage Claim~\ref{cor:Return_to_0_Proba_G} to prove Claim~\ref{lem:Transience_M} that characterizes the transience/recurrence of $M_b$ based on the comparison between~$\lambda$ and~$\lambda_{b}^{(1)}(\beta)$. 

\begin{proof}[Proof of Claim~\ref{lem:Transience_M}]
When $n_b=1$, $M_b$ is transient, because $U_b$ diverges w.p.~$1$.

The objective is to prove that $M_{b}$ is transient when $\lambda> \lambda_{b}^{(1)}(\beta)$, or equivalently, $\lambda> \lambda_{i}(\beta)$ for any item $i$ hashing to bucket $b$, i.e., $\forall i \in \cI_{b}$. Otherwise $M_{b}$ is recurrent.


Define the probability of hitting $\overline{\bm{0}}_b$, when starting in state $(i,n)$ as $r_{i,n}$, and $\bar{r}$ is the probability of hitting $\overline{\bm{0}}_b$ when starting from $\overline{\bm{0}}_b$. Formally,
\begin{align}\label{e:Return_0_From_in}
&r_{i,n} \triangleq \ProbaS{\exists t> 0:\, M_b(t)\in \overline{\bm{0}}_b}{M_b(0)=(i,n)}, \qquad i\in \cI_b, \; n\in\mathbb N\setminus \{0\} \\ \label{e:Return_0_From_0}
&\bar{r} \triangleq \ProbaS{\exists t>0:\, M_b(t) \in \overline{\mathbf 0}_b}{M_b(0)\in \overline{\mathbf 0}_b}.
\end{align}
Clearly $M_b$ is an aperiodic and irreducible Markov chain (see \eqref{e:Markov_Block1}). Thus, $M_b$ is transient, if and only if~$\bar{r}<1$, and it is recurrent if and only if~$\bar{r}=1$. From~\eqref{e:Markov_Block1}, starting at $\overline{\mathbf 0}_b$ the chain either stays at $\overline{\mathbf 0}_b$ with probability $1-\mu_{b}$ or jumps to $(i,\lambda)$ with probability $p_i$. Hence thanks to the Markov property, we can write, 
\begin{align}\label{e:R_master}
\bar{r} = (1-\mu_{b}) + \sum_{i\in \cI_{b}} p_i\,r_{i,\lambda}.
\end{align}
we deduce that $M_b$ is recurrent if and only if $r_{i,\lambda}=1$ for all $i\in \cI_{b}$, and it is transient otherwise. 

When $M_b$ is in state $(i,n)$ with $n>0$, the Markov chain $G$ defined in \eqref{e:law_Markov_G} with $p=p_i$, and $\mu=\mu_{\beta(i)}$, captures the stochastic evolution of $M_b$ until it reaches the state $(i,0)$. Thus, using Claim~\ref{cor:Return_to_0_Proba_G} we have that $r_{i,\lambda}= (r_{i,1})^{\lambda}$ with
\begin{align}
   r_{i,1}  = \begin{cases}
  1 & \text{ if }  \lambda \leq \frac{\mu_{\beta(i)}}{p_i} - 1 \\ 
   \text{Unique root in } (0,1) \text{ of } \phi(\cdot, \lambda)- \mu_{\beta(i)}/p_i, & \text{otherwise.}
   \end{cases}  
\end{align}
Noting that $\lambda_i(\beta) = \frac{\mu_{\beta(i)}}{p_i} - 1$ concludes the proof.

\end{proof}



\begin{proof}[Proof of Lemma~\ref{lem:Proba_S_infinity}]
The objective is to prove that $\Proba{S_b^{\infty}=i}=a_{i,\beta}(\lambda)$, where $a_{i,\beta}$ is defined in~\eqref{e:def_absorption_proba}.

Recall that $r_{i,n}$, defined in \eqref{e:Return_0_From_in} is the hitting probability of $M_b$ on $\overline{\bm{0}}_b$, when starting in state $(i,n)$ with $n\geq 1$, and $\bar{r}$, defined in \eqref{e:Return_0_From_0}, is the hitting probability of~$M_b$ on $\overline{\bm{0}}_b$ when starting from $\overline{\bm{0}}_b$. Both quantities were introduced in the proof of Claim~\ref{lem:Transience_M}. In particular, we observed that $r_{i,n}$ is also the hitting probability of $G$ on $0$ when starting at state $n$, when $\mu=\mu_{\beta(i)}$, and $p=p_i$. Thus, using Claim~\ref{cor:Return_to_0_Proba_G}, we deduce the expression of $r_{i,n}$, which in turn enables us to deduce $\bar{r}$ (see~\eqref{e:R_master}).

We use the notation $\ProbaC{(i,n)}{\cdot}$ to indicate the conditional probability on $M_b(0)=(i,n)$. Initially $M_b(0)\in \overline{\bm{0}}_b$. We also use the short notation $A_i^{\infty}$ to indicate the event that $S_{b}^{\infty}=i$. We compute the probability of the event $A_{i}^{\infty}$ when starting at $\overline{\bm{0}}_b$, namely $ \ProbaC{\overline{\bm{0}}_b}{A_{i}^{\infty}}$, as follows, 
\begin{align}\label{e:lem_absorption_0}
        \ProbaC{\overline{\bm{0}}_b}{A_{i}^{\infty}} 
        = 
        p_i \; \ProbaC{(i,\lambda)}{A_{i}^{\infty}}  
        +
        \sum_{j\in \cI_{b}\setminus \{i \}} p_j \; \ProbaC{(j,\lambda)}{A_{i}^{\infty}}
        +
        (1-\mu_b)\ProbaC{\overline{\bm{0}}_b}{A_{i}^{\infty}}.
\end{align}
Define $T_{\overline{\bm{0}}_b}$ as the hitting time of $M_b$ on $\overline{\bm{0}}_b$. Formally, 
\begin{align}
        T_{\overline{\bm{0}}_b} \triangleq \inf \left\{  t\geq 0: \; M_b(t) \in  \overline{\bm{0}}_b \right\}.
\end{align}
We then condition on whether $T_{\overline{\bm{0}}_b}$ is finite or infinite,

\begin{align}\nonumber 
        \ProbaC{(i,\lambda)}{A_{i}^{\infty}} 
&=
\ProbaC{(i,\lambda)}{A_{i}^{\infty}\mid T_{\overline{\bm{0}}_b}<\infty}\ProbaC{(i,\lambda)}{T_{\overline{\bm{0}}_b}<\infty} 
+ 
\ProbaC{(i,\lambda)}{A_{i}^{\infty}\mid T_{\overline{\bm{0}}_b}=\infty}\ProbaC{(i,\lambda)}{T_{\overline{\bm{0}}_b}=\infty} \\ \nonumber 
&
=
\ProbaC{\overline{\bm{0}}_b}{A_{i}^{\infty}}r_{i,\lambda} 
+
1\cdot (1-r_{i,\lambda}) \\ \label{e:lem_absorption_1}
&
=
\ProbaC{\overline{\bm{0}}_b}{A_{i}^{\infty}}r_{i,\lambda} 
+
(1-r_{i,\lambda}),
\end{align}
where we have used the strong Markov property to write that $\ProbaC{(i,\lambda)}{A_{i}^{\infty}\mid T_{\overline{\bm{0}}_b}<\infty}= \ProbaC{\overline{\bm{0}}_b}{A_{i}^{\infty}}$. Moreover, the equality $r_{i,\lambda} = \ProbaC{(i,\lambda)}{T_{\overline{\bm{0}}_b}<\infty}$ follows from the definition of $r_{i, \lambda}$ in \eqref{e:Return_0_From_in}. Similarly for $j\neq i$: 
\begin{align}\nonumber 
        \ProbaC{(j,\lambda)}{A_{i}^{\infty}} 
&=
\ProbaC{(j,\lambda)}{A_{i}^{\infty}\mid T_{\overline{\bm{0}}_b}<\infty}\ProbaC{(j,\lambda)}{T_{\overline{\bm{0}}_b}<\infty}
+ 
\ProbaC{(j,\lambda)}{A_{i}^{\infty}\mid T_{\overline{\bm{0}}_b}=\infty}\ProbaC{(j,\lambda)}{T_{\overline{\bm{0}}_b}=\infty} \\ \nonumber
&
=
\ProbaC{\overline{\bm{0}}_b}{A_{i}^{\infty}} r_{j,\lambda} +
0 \cdot (1-r_{j,\lambda}) \\ \label{e:lem_absorption_2}
&
= 
\ProbaC{\overline{\bm{0}}_b}{A_{i}^{\infty}} r_{j,\lambda}.
\end{align}

\noindent Combining \eqref{e:lem_absorption_0}, \eqref{e:lem_absorption_1}, and \eqref{e:lem_absorption_2}, we deduce that, 
\begin{align}
\ProbaC{\overline{\bm{0}}_b}{A_{i}^{\infty}}
= 
\frac{p_i(1-r_{i,\lambda})}{\mu_b-\sum_{j\in \cI_{b}} p_jr_{j,\lambda}} 
=  
\frac{p_i(1-r_{i,\lambda})}{\sum_{j\in \cI_{b}} p_j(1-r_{j,\lambda})}. 
\end{align}
Observe that $r_{i,\lambda}=(r_{i,1})^{\lambda}$, and $r(\nicefrac{\mu_{\beta(i)}}{p_i},\lambda)= r_{i,1}$, where the function $r(\cdot, \cdot)$ is defined in~\eqref{e:def_r_root_function}. Thus, $\ProbaC{\overline{\bm{0}}_b}{A_{i}^{\infty}}= a_{i,\beta}(\lambda)$, which finishes the proof.


\end{proof}


\subsection{Proof of items~\ref{item:thm_limit_V} and~\ref{item:thm_limit_Y} of Theorem~\ref{thm:counters_limit}}
\label{ss:proof_limit_V_and_Y}

We prove the limiting distribution for $V_{b}^{+}$ and $Y_{l,c}$ in equations~\eqref{e:limit_V_plus} and~\eqref{e:limit_y}. We derive in Lemma~\ref{lem:Count_ElasticSketch_Finite} a finite time characterization of the counters $V_b^{+}$ and $Y_{\ell,c}$ in terms of the counts vector $\bm{N}$ and the last eviction time $T_{b}^{\text{evict}}$.

\begin{lemmaSp}\label{lem:Count_ElasticSketch_Finite}
The following holds at any step $t$,  
\begin{align}\label{e:Vplus_eq}
&V_b^{+}(t) = N_{S_b(t)}(t) - N_{S_b(t)}(T_b^{\text{evict}}(t)), \\ \label{e:YCMS_eq}
&Y_{\ell,c}(t) = Y_{\ell,c}^{\CM}(t) - \sum_{b=1}^{m_1} \left(N_{S_b(t)}(t) - N_{S_b(t)}(T_{b}^{\text{evict}}(t)) \right) \mathds{1}\left( h_{\ell}(S_b(t)) =c\right),
\end{align}
\end{lemmaSp}

The proof of Lemma~\ref{lem:Count_ElasticSketch_Finite} is presented in Section~\ref{app:proof_lemma_finite_time_counters} of the appendix.

Combining Lemmas~\ref{lem:Relation_S_M} and~\ref{lem:Count_ElasticSketch_Finite}, and Claim~\ref{lem:Transience_M}, for any $b^{+}\in \cB_{\beta}^{+}(\lambda)$ as $t\to\infty$, the following holds a.s., 
   \begin{align}\label{e:bplus_formula}
    \lim_{t\to \infty} \frac{1}{t}\left(N_{S_{b^{+}}(t)}(t) - N_{S_{b^{+}}(t)}(T^{\text{evict}}_{b^+}(t)) \right) =  p_{S_{b^{+}}(\infty)} -   \lim_{t\to \infty} \frac{N_{S_{b^+}(t)}(T^{\text{evict}}_{b^{+}}(t))}{t} =  p_{S_{b^+}^{\infty}},
   \end{align}
where we have used the fact that $\lim_{t\to \infty} N_i(t)/t=p_i$, and $N_{i}(T_b^{\text{evict}}(t))\leq T_{b}^{\text{evict}}$ for any item $i$. On the other hand, for any $b^{-}\in \cB_{\beta}^{-}(\lambda)$, we have that a.s., 
\begin{align}\label{e:bminus_formula}
          \lim_{t\to \infty}    \frac{1}{t}  \left(N_{S_{b^{-}}(t)}(t) - N_{S_{b^{-}}(t)}(T^{\text{evict}}_{b^{-}}(t)) \right)  \leq \lim_{t\to \infty}  \frac{t-T^{\text{evict}}_{b^{-}}(t)}{t} = 0,
\end{align} 
where we have used the fact that $\left(N_{i}(t) - N_i(T^{\text{evict}}_{b^{-}}(t)) \right)$ is the number of appearances of item $i$ between time steps $t$ and $T^{\text{evict}}_{b^{-}}(t)+1$. Plugging~\eqref{e:bplus_formula} and~\eqref{e:bminus_formula} in Lemma~\ref{lem:Count_ElasticSketch_Finite}, we prove \eqref{e:limit_V_plus} and \eqref{e:limit_y}.









\section{Conclusion}
\label{s:Conclusion}

We studied the limiting behavior of~$\ElaSk$ under a random stationary stream model and derived closed-form expressions for the asymptotic expected counting error. These expressions enable an efficient grid search for the near-optimal configuration of the eviction threshold~$\lambda$ and the memory split between the heavy and~$\CM$ blocks controlled via $m_1$. Moreover, our characterization of the optimal threshold restricts the search for~$\lambda$ to a small candidate set, substantially reducing its computational cost. In future work, we aim to move beyond grid search by deriving explicit tuning rules for $(\lambda,m_1)$ as a function of the arrival distribution~$\bm p$.



\appendix 

\section{Proof of Lemma~\ref{lem:Count_ElasticSketch_Finite}}
\label{app:proof_lemma_finite_time_counters}
\begin{proof}[Proof of Lemma~\ref{lem:Count_ElasticSketch_Finite}]

We first observe that \eqref{e:YCMS_eq} is equivalent to, 
\begin{align}\label{e:YElastic_Form2}
 Y_{\ell,c}(t) = \sum_{i\in \cI} \delta_{\ell,c}(i) \left[ (1-\delta_i(t,\beta)) N_i(t) + \delta_i(t,\beta)N_i(T^{\text{evict}}_{\beta(i)}(t)) \right], \; \delta_i(t,\beta) = \mathds{1}\left( S_{\beta(i)}(t) = i \right),
\end{align}
where $\delta_{\ell,c}(i)=\mathds{1}(h_l(i)=c)$.
 Since $Y_{\ell,c}^{\CM}(t)=\sum_{i\in \mathcal{I}} \delta_{\ell,c}(i) N_i(t)$, the difference $Y_{\ell,c}^{\CM}(t)-Y_{\ell,c}(t)$ from \eqref{e:YElastic_Form2} can be partitioned into $m_1$ terms each corresponding to items in $\cI_b$. In each one of these partitions, $\delta_i(t,\beta)$ is only equal to $1$ for $S_b(t)$ and hence \eqref{e:YCMS_eq} is equivalent to~\eqref{e:YElastic_Form2}. The proof of \eqref{e:YElastic_Form2} follows an induction argument.

\noindent\textbf{Base case $t=0$.}\;
All vectors are $0$ (or $-1$), hence both sides of \eqref{e:Vplus_eq} and \eqref{e:YCMS_eq} vanish.

\smallskip
\noindent\textbf{Induction step.}
Let $b=\beta(R_t)$ and assume \eqref{e:YCMS_eq} and \eqref{e:Vplus_eq} hold at step $t-1$. 
We show they still hold after processing $R_t$.

\smallskip
\noindent\textbf{Case 1: $S_b(t-1)=-1$ (empty bucket).}\;
The algorithm stores $S_b(t)=R_t$, sets $V_b^{+}(t)=1$, leaves $\bm{Y}$ unchanged, and $T_b^{\text{evict}}(t)=t-1$.
Because $N_{R_t}(t)=N_{R_t}(t-1)+1$, the right–hand side of \eqref{e:Vplus_eq} equals
$N_{R_t}(t)-N_{R_t}(t-1)=1=V_b^{+}(t)$.
For \eqref{e:YCMS_eq}, the item $R_t$ switches from unmonitored to monitored ($\delta_{R_t}(t,\beta)=1$ and $\delta_{R_t}(t-1,\beta)=0$)
its contribution changes from $N_{R_t}(t-1)$ to
$N_{R_t}(T_b^{\text{evict}}(t))=N_{R_t}(t-1)$. Hence the sum is unchanged, matching the unchanged $\bm{Y}$.

\smallskip
\noindent\textbf{Case 2: $S_b(t-1)=R_t$ (bucket hit).} 
Here $S_b(t-1)=S_b(t)$, $T_b^{\text{evict}}(t)= T_b^{\text{evict}}(t-1)$, and $V_{b}^{+}(t)= V_b^+(t-1)+1$. Thus \eqref{e:Vplus_eq} is preserved because $N_{R_t}(t)-N_{R_t}(T_b^{\text{evict}}(t))=N_{R_t}(t-1)+1-N_{R_t}(T_b^{\text{evict}}(t-1))$.
$\bm{Y}$ is untouched and $R_t$ remains monitored, so its term in
\eqref{e:YCMS_eq} is still $N_{R_t}(T_b^{\text{evict}}(t-1))$. Therefore, the sum does not
change and \eqref{e:YCMS_eq} holds.

\smallskip
\noindent\textbf{Case 3: $S_b(t-1) \neq R_t$ and $\lambda V_b^{+}(t-1) - V_b^{-}(t-1) > 0$ (mismatch, vote passes).}\;
The bucket content is not modified ($S_b(t-1)=S_b(t)$, $T_b^{\text{evict}}(t-1)=T_b^{\text{evict}}(t)$, and $V_b^{+}(t)= V_{b}^{+}(t-1)$), only $V_b^{-}$ increases, and $\bm{Y}$ is updated by incrementing the cells corresponding to $R_t$ by $1$. Thus both sides of equation~\eqref{e:Vplus_eq} remains unchanged. Moreover, $\delta_{R_t}(t,\beta)=\delta_{R_t}(t-1,\beta)=0$ and hence the RHS of~\eqref{e:YCMS_eq} for all cells $(\ell,h_{\ell}(R_t))$ with $r\in [d]$ increase by $1$ and thus \eqref{e:YCMS_eq} holds at step $t$.

\smallskip

 \noindent\textbf{Case 4: $S_b(t-1)\neq R_t$ and $\lambda V_b^{+}(t-1)-V_b^{-}(t-1)\le0$ (mismatch, vote fails, eviction).}  
Let $j=S_b(t-1)$.  The algorithm evicts $j$, adds $V_b^{+}(t-1)$ to every CMS cell $(\ell,h_{\ell}(j))$, then sets
$S_b(t)=R_t$, $V_b^{+}(t)=1$, $V_b^{-}(t)=0$, and $T_b^{\text{evict}}(t)=t-1$. Since $N_{R_t}(t)=N_{R_t}(t-1)+1$, the right–hand side of \eqref{e:Vplus_eq} becomes $N_{R_t}(t)-N_{R_t}(t-1)=1=V_b^{+}(t)$, so \eqref{e:Vplus_eq} holds. For \eqref{e:YCMS_eq}, note that 1) For any cell $(\ell,c)\neq(\ell,h_{\ell}(j))$, the term for $R_t$ changes by  
  $\delta_{\ell,c}(R_t)\left(N_{R_t}(T_b^{\text{evict}}(t))-N_{R_t}(t-1)\right)=0$,  
  and no other contributions change, so $Y_{\ell,c}(t)=Y_{\ell,c}(t-1)$, 2) For each cell $(\ell,h_{\ell}(j))$, the term for $j$ increases by  
  $N_j(t)-N_j(T_b^{\text{evict}}(t-1))=V_b^{+}(t-1)$, matching the $\CM$ update. Hence \eqref{e:YCMS_eq} holds at step $t$ in all cells.


\smallskip
\noindent All cases have been tested, which finishes the proof. 

\end{proof}

\section{Proof of Lemma~\ref{lem:g_b_decreasing}}
\label{app:proof_thm_lambda_star}


For lighter notation, we drop the subscript referring to the hash function~$\beta$. The objective is to prove that $g_b$ is decreasing over $(\lambda_b^{(1)},\infty)$. To this end, we prove that for any $\ell\geq 2$,
    \begin{enumerate}

         \item The left limit of~$g_b$ at~$\lambda_b^{(\ell)}$ is larger than the right limit at the same point, i.e.,
         \begin{align}\label{e:g_b_continuous_decreasing}
                 \lim_{\lambda\to\lambda^{(\ell)}-} g_b(\lambda) \geq \lim_{\lambda\to\lambda^{(\ell)}+} g_b(\lambda) 
         \end{align}

        \item The function $g_b$ is decreasing in the interval~$(\lambda_b^{(\ell)}, \lambda_b^{(\ell+1)}]$.
    \end{enumerate}
       
\medskip 

\noindent \textbf{Proof of~\eqref{e:g_b_continuous_decreasing}.} For brevity write
$\lambda^{(\ell)} = \lambda_b^{(\ell)}$,
$p^{(j)} =  p_{b}^{(j)}$,
$w^{(j)} = p^{(j)}  w \left(\lambda^{(\ell)}, \mu_{b}/p^{(j)}\right)$,
$u_{\ell} = \sum_{j=1}^{\ell-1} w^{(j)}$,
and $ v_{\ell} = \sum_{j=1}^{\ell-1} p^{(j)} w^{(j)}$.
For $\lambda < \lambda^{(\ell)}$ there are~$\ell-1$ active items (any item $i$ such that $a_i(\lambda)>0$) in bucket $b$, and for $\lambda > \lambda^{(\ell)}$ there are $\ell$ active items. Therefore the left and right limits of $g_b$ at $\lambda^{(\ell)}$ can be written as, 
\begin{align}
  \lim_{\lambda\to\lambda^{(\ell)}-} g_b(\lambda)
  = \frac{v_{\ell}}{u_{\ell}} , \; \;  \lim_{\lambda\to\lambda^{(\ell)}+} g_b(\lambda)
  = \frac{v_{\ell} + p^{(\ell)} w^{(\ell)}}{u_{\ell} + w^{(\ell)}} .
\end{align}
Therefore,
\begin{align}
  \lim_{\lambda\to\lambda^{(\ell)}-} g_b(\lambda) - \lim_{\lambda\to\lambda^{(\ell)}+} g_b(\lambda) 
= \frac{w^{(\ell)} v_{\ell} - p^{(\ell)} w^{(\ell)} u_{\ell}}{u_{\ell}\left(u_{\ell} + w^{(\ell)}\right)}
= \frac{w^{(\ell)}}{u_{\ell}\left(u_{\ell} + w^{(\ell)}\right)} \sum_{j=1}^{\ell-1} w^{(j)} \left(p^{(j)} - p^{(\ell)}\right) > 0.
\end{align}

\medskip

\noindent \textbf{Proof that $g_b$ is decreasing over $(\lambda_b^{(\ell)},\lambda_b^{(\ell+1)}]$.} Define the function $\zeta:[1,+\infty)^2 \mapsto \mathbb{R}$ as, 
\begin{align}\label{e:def_zeta}
  \zeta(\lambda,C) \triangleq \frac{\partial }{\partial \lambda} \ln w(\lambda,C).
\end{align}

\noindent 

We prove in Lemma~\ref{lem:sign_gb_prime} that if~$\zeta(\lambda,\cdot)$ is increasing, then~$g_b$ is decreasing over~$(\lambda_b^{(\ell)},\lambda_b^{(\ell+1)}]$, and we prove in Lemma~\ref{lem:increasing_zeta} that $\zeta(\lambda,\cdot)$ is indeed increasing, to deduce that $g_b$ is decreasing over $(\lambda_b^{(\ell)},\lambda_b^{(\ell+1)}]$. The two lemmas are presented below along with their proofs.

\begin{lemmaSp}\label{lem:sign_gb_prime}
    Let~$\lambda \in (\lambda_b^{(\ell)},\lambda_b^{(l+1)}]$ for some~$l$. If~$\zeta(\lambda, \cdot)$ is increasing then~$g_b$ is decreasing in~$\lambda$. 
\end{lemmaSp}

\begin{proof}[Proof of Lemma~\ref{lem:sign_gb_prime}]

Here, we drop the subscripts that indicate the choice of the hash function $\beta$ for lighter notation. 
Using the definitions of the functions in~\eqref{e:def_absorption_weights},~\eqref{e:def_absorption_proba},  and~\eqref{e:def_g_and_gb}, $g_b$ can be written as,  
    \begin{align}
           g_b(\lambda) =\frac{\sum_{j\in \mathcal{A}}  p_j w_{j}(\lambda)  }{ \sum_{k\in \mathcal{A}} w_{k}(\lambda)},
    \end{align}    
where~$\cA$ is the set of active items in bucket~$b$, i.e., $\cA \triangleq \{i \in \beta_b^{-1}: \; \lambda_i < \lambda_b^{(l+1)} \}$. Further define the sets $\cA^{2,+}$ and $\cA^{2,-}$ as,
\begin{align}
    \cA^{2,+} \triangleq \{ (j,k)\in \cA^{2}: p_{j}> p_k \}, \; \cA^{2,-} \triangleq \{ (j,k)\in \cA^{2}: p_{j}< p_k \},
\end{align}
and $C_j = P_{\beta(j)}/p_j$. 
The sign of the derivative of $g_b$ matches the sign of $\gamma$, that is given by, 
    \begin{align}
           \gamma 
&=  \sum_{(j,k)\in \mathcal{A}^{2}} p_j w_{j}^{'}(\lambda) w_k(\lambda)  -  \sum_{(j,k)\in \mathcal{A}^{2}} p_j w_{j}(\lambda) w_k^{'}(\lambda) \\ 
&= \sum_{(j,k)\in \mathcal{A}^{2}} p_j w_{j}^{'}(\lambda) w_k(\lambda) - \sum_{(j,k)\in \mathcal{A}^{2}} p_k w_{k}(\lambda) w_j^{'}(\lambda) \\ 
&= \sum_{(j,k)\in \mathcal{A}^{2}} w_k(\lambda) w_j^{'}(\lambda) \left(p_j - p_k\right) \\ 
& = \sum_{(j,k)\in \cA^{2,+}} (p_j - p_k)  w_k(\lambda) w_j^{'}(\lambda) + \sum_{(j,k)\in \cA^{2,-}} (p_j-p_k)  w_k(\lambda) w_j^{'}(\lambda) \\ 
& = \sum_{(j,k)\in \cA^{2,+}} (p_j - p_k)  w_k(\lambda) w_j^{'}(\lambda) + \sum_{(j,k)\in \cA^{2,+}} (p_k-p_j)  w_j(\lambda) w_k^{'}(\lambda) \\ 
&= \sum_{(j,k)\in \cA^{2,+}}   (p_j - p_k) \left(  w_k(\lambda)w_j^{'}(\lambda) - w_j(\lambda) w_k^{'}(\lambda) \right) \\ 
&= \sum_{(j,k)\in \cA^{2,+}}(p_j - p_k)p_k p_j \left( w(\lambda, C_k) \partial_{\lambda} w(\lambda, C_j) - w(\lambda, C_j) \partial_{\lambda} w(\lambda, C_k)  \right) \\
&= \sum_{(j,k)\in \cA^{2,+}}(p_j-p_k)p_kp_j \left(w(\lambda, C_j) w(\lambda, C_k) \right) \left( \zeta(\lambda, C_j) - \zeta(\lambda, C_k) \right).
    \end{align}
Observe that any two items~$j$ and~$k$ in~$\cA^{2,+}$ hash to the same bucket with $p_j>p_k$, and thus $C_j<C_k$. Therefore, if $\zeta(\lambda, \cdot)$ is increasing, then $g_b$ is decreasing, which finishes the proof.  
\end{proof}


\begin{lemmaSp}\label{lem:increasing_zeta}  
For any $\lambda\geq 1$, the function $\zeta(\lambda, \cdot)$, defined in \eqref{e:def_zeta}, is increasing. 
\end{lemmaSp}

\begin{proof}

We first derive the partial derivatives of the function $r$, defined in~\eqref{e:def_r_root_function}, and deduce that $r$ is decreasing in $\lambda$ and increasing in $C$. We then leverage these expressions to deduce that~$\zeta$ can be written as,
\begin{align}\label{e:def_function_f}
  \zeta(\lambda,C) = f(r(\lambda, C)): \; f(x) \triangleq  \frac{-x^{\lambda} \ln x}{\lambda x^{\lambda + 1} -(\lambda+1) x^{\lambda} + 1 }.
\end{align}
Next, we prove that  $f$ is increasing over $(0,1)$, and given that $r$ is increasing in $C$, we deduce that $\zeta$ is increasing in $C$.

\medskip 

\noindent \textbf{Partial Derivatives of $r$.} For brevity write $r = r(\lambda,C)$ and assume that $\lambda> C-1$ Let $\partial_{\lambda} r$ and~$\partial_{C} r$ designate the partial derivatives of~$r$ (see~\eqref{e:def_r_root_function}) with respect to~$\lambda$ and~$C$, respectively. We prove that, 
  \begin{align}\label{e:derivative_r_formula}
        \partial_{\lambda} r =  \frac{(r^{\lambda+1}\ln r) (1-r)}{\lambda r^{\lambda+1}-(\lambda+1)r^{\lambda}+1},\; \; 
        \partial_{C} r =  \frac{(r-1)^{2}}{\lambda r^{\lambda+1}-(\lambda+1)r^{\lambda}+1}. 
\end{align}
 We differentiate the equation~$\phi(r,\lambda)= C$ to deduce that,
\begin{align}\label{e:partial_derivatives_r}
     \partial_{\lambda} r = -\frac{\partial_2 \phi(r,\lambda)}{\partial_1 \phi(r,\lambda)}, \; \;
     \partial_{C} r = \frac{1}{\partial_1 \phi(r,\lambda)}, 
\end{align}
where $\partial_1 \phi$ and $\partial_2 \phi$ designate the partial derivatives of $\phi$ with respect to the first and second components, respectively. Direct calculation yields, 
\begin{align}
  \partial_{2} \phi(r,\lambda)= \frac{r^{\lambda+1}\ln r}{r-1}, \; \; 
  \partial_{1} \phi(r,\lambda), \frac{\lambda r^{\lambda+1}-(\lambda+1)r^{\lambda}+1}{(r-1)^2}, 
\end{align}
which enables us to deduce~\eqref{e:derivative_r_formula}. Note that $x\mapsto\lambda x^{\lambda+1}-(\lambda+1)x^{\lambda}+1$ is decreasing over $(0,1)$---direct computation of the derivative---and its value at $0$ and $1$ are positive, and thus $ \partial_{\lambda} r\leq 0$ and $\partial_{C} r \geq 0$. 

\medskip 

\noindent \textbf{Proof of \eqref{e:def_function_f}.} From $\phi(r,\lambda)=C$ we get,
\begin{align}\label{e:w_simplification}
C-1= \frac{r-r^{\lambda+1}}{1-r} =  \frac{rw(\lambda,C)}{1-r} \implies  w(\lambda,C) = (C-1) \frac{1-r}{r} \implies \zeta(\lambda,C) =  \frac{\partial_\lambda r}{r(1-r)}.
\end{align}
Using the expression in~\eqref{e:partial_derivatives_r}, we deduce \eqref{e:def_function_f}. 

\medskip 

\noindent \textbf{Proof that $f$ is increasing over $(0,1)$.} Let $z=-\ln x\in(0,\infty)$ so that $x=e^{-z}$. Then
\begin{align}
f\left(e^{-z}\right)
&=\frac{-e^{-\lambda z}\ln\left(e^{-z}\right)}{\lambda e^{-(\lambda+1)z}-(\lambda+1)e^{-\lambda z}+1}
=\frac{z}{\alpha(z)},
\\
\alpha(z) &\triangleq e^{\lambda z}-(\lambda+1)+\lambda e^{-z}.
\end{align}
We have $\alpha(0)=0$ and for $z>0$, $\alpha'(z)=\lambda\left(e^{\lambda z}-e^{-z}\right)>0$, so $\alpha(z)>0$. Moreover, $\alpha''(z)=\lambda^2 e^{\lambda z}+\lambda e^{-z}>0$, hence $\alpha$ is convex on $(0,\infty)$. Fix $0<u<z$ and set $\theta=u/z\in(0,1)$. By convexity,
\begin{align}
\alpha(u)=\alpha\left(\theta z+\left(1-\theta\right)0\right)\le \theta \alpha(z)+\left(1-\theta\right)\alpha(0)=\frac{u}{z}\alpha(z),
\implies \frac{z}{\alpha(z)}\le \frac{u}{\alpha(u)}.
\end{align}
Thus $z\mapsto f\left(e^{-z}\right)=z/\alpha(z)$ is decreasing on $(0,\infty)$. Since $z=-\ln x$ is decreasing in $x$ on $(0,1)$, it follows that $x\mapsto f(x)$ is increasing on $(0,1)$. This finishes the proof.

\end{proof}

\section{Probabilistic Bounds for the Optimal Eviction Threshold}
\label{app:proba_bounds_lambda_star}
The high-probability bound on $\lambda^{*}(\beta)$ in Theorem~\ref{thm:hp_lambda} is distribution-agnostic and therefore does not capture how the skew of the arrival distribution~$\bm p$ affects~$\lambda^{*}(\beta)$. A more $\bm p$-sensitive route is to upper-bound $\max_{b\in\cB}\lambda_b^{(1)}(\beta)$ and then apply a union bound over buckets.

Assume for simplicity that $\cI=[n_{\cI}]$ and $p_1>p_2>\cdots>p_{n_{\cI}}$. Fix a bucket $b\in\cB$ and condition on its \textit{dominant} item, i.e., the highest-probability item among those hashing to~$b$. The probability that the dominant item in bucket $b$ is item $j$ occurs with probability
$
\frac{1}{m_1}\left(1-\frac{1}{m_1}\right)^{j-1}
$
.Conditioned on this event (so that $p_b^{(1)}=p_j$), $\lambda_{b}^{(1)}$ can be written as,
\begin{equation}\label{e:lambda_b1_cond}
\lambda_b^{(1)}(\beta)
=\frac{\mu_b}{p_j}-1
=\sum_{k=j+1}^{n_{\cI}} \mathds{1}\{\beta(k)=b\}\,\frac{p_k}{p_j}.
\end{equation}
The right-hand side is a weighted sum of independent Bernoulli indicators (each $\mathds{1}\{\beta(k)=b\}$ has mean $1/m_1$). Thus, for each fixed $j$, one can apply Bernstein's inequality to control $\lambda_b^{(1)}(\beta)$ under the conditioning. Averaging over $j$ and union-bounding over buckets yields a $\bm p$-dependent high-probability upper bound on $\lambda^{*}(\beta)$.





\end{document}